\documentclass[letterpaper]{article} 
\usepackage[draft]{aaai2026}  
\usepackage{times}  
\usepackage{helvet}  
\usepackage{courier}  
\usepackage[hyphens]{url}  
\usepackage{graphicx} 
\usepackage{natbib}  
\usepackage{caption} 
\usepackage{algorithm}
\usepackage{algorithmic}

\usepackage{subcaption}
\usepackage{multirow}
\usepackage{booktabs}

\usepackage{newfloat}
\usepackage{listings}
\DeclareCaptionStyle{ruled}{labelfont=normalfont,labelsep=colon,strut=off} 
\floatstyle{ruled}
\newfloat{listing}{tb}{lst}{}
\floatname{listing}{Listing}
\title{An AI-Powered Framework for Analyzing Collective Idea Evolution in Deliberative Assemblies}
\author{
    Elinor Poole-Dayan,
    Deb Roy, and 
    Jad Kabbara
}
\affiliations{
    Center for Constructive Communication\\
    Massachusetts Institute of Technology\\
    elinorpd@mit.edu
}

\usepackage{microtype}
\usepackage{url}
\usepackage{booktabs}
\usepackage{enumitem}
\usepackage{mdframed}
\usepackage{listings}
\usepackage[nameinlink]{cleveref}

\begin{document}

\maketitle

\begin{abstract}
In an era of increasing societal fragmentation, political polarization, and erosion of public trust in institutions, representative deliberative assemblies are emerging as a promising democratic forum for developing effective policy outcomes on complex global issues. Despite theoretical attention, there remains limited empirical work that systematically traces how specific ideas evolve, are prioritized, or are discarded during deliberation to form policy recommendations. Addressing these gaps, this work poses two central questions: (1) How might we trace the evolution and distillation of ideas into concrete recommendations within deliberative assemblies? (2) How does the deliberative process shape delegate perspectives and influence voting dynamics over the course of the assembly? To address these questions, we develop LLM-based methodologies for empirically analyzing transcripts from a tech-enhanced in-person deliberative assembly. The framework identifies and visualizes the space of expressed suggestions. We also empirically reconstruct each delegate's evolving perspective throughout the assembly. 
Our methods contribute novel empirical insights into deliberative processes and demonstrate how LLMs can surface high-resolution dynamics otherwise invisible in traditional assembly outputs.
\end{abstract}


\section{Introduction}

In an era of increasing societal fragmentation, political polarization, and erosion of public trust in institutions, representative \textit{deliberative assemblies}\footnote{Also referred to as a Citizens’ Assembly, Civic Assembly, or Citizens’ Jury.} are emerging as a promising democratic forum for developing effective policy outcomes on complex global issues \citep{oecd_open_2016, oecd_innovative_2020, dryzek_crisis_2019}. 

A deliberative assembly consists of a representative sample of citizens—called delegates—selected by lottery. They gather over several days to find common ground and form collective \textit{recommendations} for decision-makers through collaborative learning and facilitated deliberation \citep{cesnulaityte_assembling_2023}. Deliberative assemblies have been strikingly successful in tackling complex policy problems and values-based dilemmas, from constitutional changes around same-sex marriage and abortion to urban planning, clean energy, and climate change \citep{oecd_innovative_2020}. Deliberative assemblies facilitate the rebuilding of trust necessary for effective, fair, and legitimate democratic decision-making by fostering meaningful dialogue, collaborative learning, and bridging diverse viewpoints \citep{oecd_innovative_2020}.

While promising, deliberative assemblies present logistical challenges, are difficult to design and facilitate effectively, and are virtually impossible to scale up \citep{dryzek_crisis_2019}. Moreover, the theoretical foundation of deliberative democracy emphasizes the role of informed and meaningful discussion in legitimizing decision-making processes \citep{habermas_theory_1984}. 
Despite theoretical attention, there remains limited empirical work that systematically traces how specific ideas evolve, are prioritized, or are discarded during deliberation to form policy recommendations.

Addressing these gaps, this work poses two central questions. 

First, \textbf{how might we trace the evolution and distillation of ideas into concrete recommendations within deliberative assemblies?}
Recent advances in large language models (LLMs) have shown unprecedented synthesis and reasoning capabilities, especially when combined with prompting methods such as chain-of-thought \citep{wei_chain--thought_2022}.
In the novel context of an in-person deliberative assembly, we leverage LLMs to make several key empirical contributions to understanding the ways in which deliberation evokes, shapes, and distills a wealth of ideas into actionable recommendations.

\begin{enumerate}
    \item First, we design an LLM-based methodological framework to operationalize, identify, visualize, and analyze idea evolution directly from the transcripts of recorded deliberative assemblies. 
    \item Our analyses reveal seemingly large discrepancies between the ideas expressed during the assembly and the final recommendations. However, we demonstrate that this often, in fact, signals healthy deliberative processes—where ideas were surfaced, refined, or appropriately discarded. 
    \item Leveraging our framework, we recover valuable ideas that were organically missed during the assembly, laying the groundwork for transforming this framework into a tool for live integration in future deliberative assemblies.
\end{enumerate}

Second, \textbf{how does the deliberative process shape delegate perspectives and influence voting dynamics over the course of the assembly?}

We address this with the following contributions: 
\begin{enumerate}
 \item {Delegate profiles}: We develop an LLM‐based framework for reconstructing each delegate’s evolving perspective across the stages of deliberation via compact, modular representations empirically grounded in the transcripts. 
    
    \item Leveraging the profiles, we uncover \emph{when} and \emph{why} delegates revised their views, anchoring the drivers of vote change to the exact moments they emerged in the deliberative process. 
    
    \item We demonstrate how the profiles reveal underlying values, trade-offs, and transformations that shaped final outcomes, enabling novel insights that can be shared post-assembly to increase accountability and visibility into the process.

    \end{enumerate}

While grounded in this specific use case, the framework is designed to be adaptable to future assemblies, offering a foundation for exploring how deliberative processes can vary across settings and issue domains. The empirical insights we uncover shed new light on idea evolution and decision-making in deliberative contexts, contributing to a growing understanding of how these processes unfold in practice.

\section{Background \& Related Work} 
Deliberative democracy is a theoretical framework that emphasizes the importance of informed and meaningful discussion in legitimate decision-making processes, challenging traditional democratic practices that prioritize mere vote aggregation \cite{habermas_theory_1984,dryzek_crisis_2019}. This approach advocates for rigorous public discourse where diverse perspectives are considered to achieve reasoned consensus, thereby enhancing the legitimacy of democratic outcomes. Placing citizens at the heart of democratic power is not only seen as more legitimate, but is increasingly viewed as an urgent and necessary response to counteract the breakdown of trust between the public and governing institutions dominated by elites 
\cite{landemore_open_2020,oecd_open_2016,oecd_innovative_2020}.

Deliberative theory has offered extensive normative accounts of how public reasoning ought to function—emphasizing inclusion, mutual justification, and rational consensus \cite{habermas_theory_1984,seidman_contributions_2020}. 
Scholars such as \citet{fishkin_when_2011} and \citet{mansbridge_systemic_2012} have highlighted that real-world deliberation often departs from idealized conditions, with uneven participation, strategic framing, or opinion shifts occurring over time. Yet, much of this work relies on observational or survey-based methods. 
There is a need for empirical understanding of how core dynamics—like surfacing, refining, and prioritizing suggestions; evaluating feasibility and trade-offs; and facilitation practices—unfold in real assemblies. This lack of transparency risks undermining the legitimacy and trust in democratic deliberation, especially as decision-makers often receive final recommendations without insights into the deliberative processes behind them \citep{fishkin_when_2011, mansbridge_systemic_2012, kundnani_future_2020, oecd_innovative_2020}.

Prior work on technology for deliberation largely focuses on building online deliberation platforms \citep{small_polis_2021,konya_democratic_2023,fishkin_deliberative_2019}, automating facilitation of online deliberative discussions \citep{argyle_leveraging_2023,hadfi_conversational_2023,wyss_artificial_2017,fishkin_deliberative_2019,shin_chatbots_2022, kim_bot_2020}, and leveraging LLMs for consensus-finding \citep{bakker_fine-tuning_2022,jarrett_language_2023,tessler_ai_2024,fish_generative_2023,small_opportunities_2023}. 
In contrast, our work is the first to analyze the fine-grained evolution of ideas and reasoning over the course of an \textit{in-person} deliberative assembly. By leveraging full transcripts and LLM-based methods, our approach enables a new type of analysis that reveals how suggestions emerge, transform, and influence delegate perspectives in real-time. This work thus adds a complementary empirical lens to the deliberative technology landscape, particularly suited to understanding the internal dynamics of assemblies.

\section{Deliberative Assembly}\label{sec:data}
The core contributions of this work are uniquely grounded in a real deliberative assembly. There are no existing datasets of deliberative assemblies, so we designed and hosted a three-day deliberative assembly composed of nineteen undergraduate student delegates at MIT. Beyond data collection, this assembly served as an authentic opportunity for undergraduate students to deliberate on sustainability policy, where their goal was ``\textit{generating actionable recommendations that position MIT as a leader in campus sustainability and serve as a model for other institutions}."
Ultimately, the delegates authored and passed 16 recommendations which were delivered to the MIT Office of Sustainability.\footnote{For more info on the assembly, insights, and outcomes, see \url{https://sustainabilityassembly.portal.cortico.ai/}}

\subsection{Assembly Design}
\begin{figure}
    \centering
    \includegraphics[width=\linewidth]{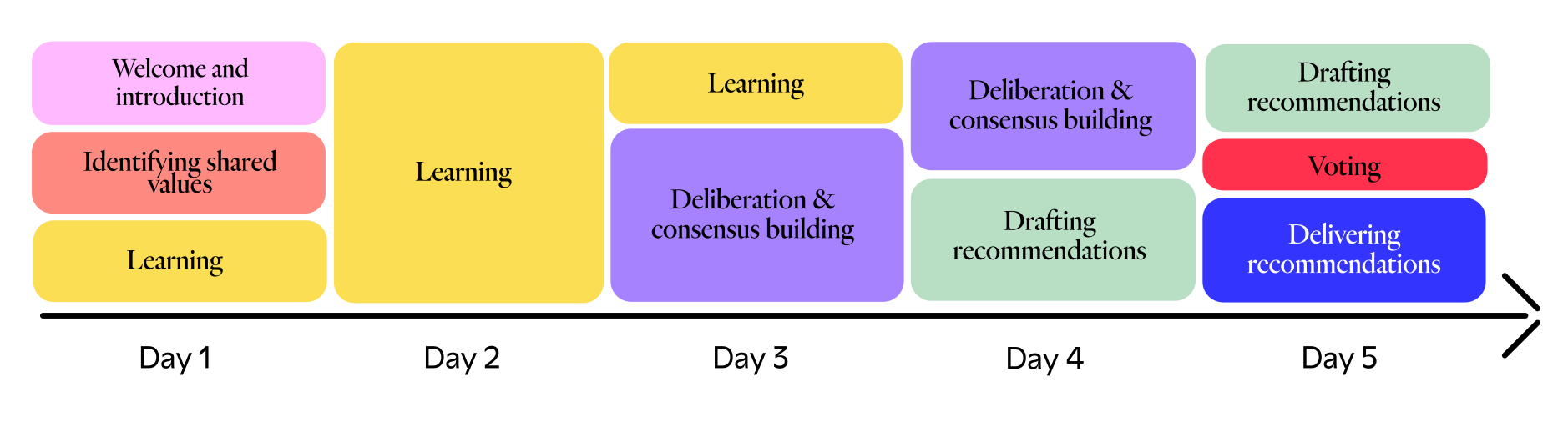}
    \caption{Example breakdown of deliberative assembly phases. Graphic courtesy of DemocracyNext \citep{cesnulaityte_assembling_2023}.}
    \label{fig:phases}
\end{figure}
The assembly itself took place over three days with a six-day gap between the second and last days. The structure followed that of a traditional deliberative assembly (\Cref{fig:phases}), each day consisting of a series of moderated conversations and activities, some involving the full group (``plenaries") and others in smaller groups of 4-5 students (``breakouts") with one staff member moderating the discussion.

\paragraph{Day 1.}
The first day began with introductions, an overview of the schedule, and creating ground rules. The assembly began in the learning phase, where delegates shared personal experiences, discussed readings, and identified gaps in their knowledge to prepare questions for sustainability experts. 

\paragraph{Day 2.}
The learning phase continued on the second day with a plenary panel discussion with four on-campus sustainability experts. It then transitioned to the deliberation phase where delegates debriefed the expert panel in breakout groups and then started drafting recommendations on paper. They participated in a gallery walk in which they could view all of the written recommendations, leave comments, and group redundant recommendations together. These were then transcribed into a spreadsheet and voted upon. This preliminary vote served to give the delegates a sense of which recommendations needed additional revisions or where questions remained.

\paragraph{Day 3.}
The third day continued the deliberation and recommending phases of the assembly. In the ``Information Fair," delegates had another opportunity to meet with seven sustainability experts to ask any unresolved questions and consult them on their drafted recommendations. Then, the delegates debriefed their takeaways in breakout groups and revised their written recommendations. Finally, the delegates voted on the final set of revised recommendations, those that received a supermajority (at least 80\%) approval became the final set of assembly recommendations. 
Sixteen final policy recommendations passed out of 29 drafted recommendations and were delivered to the Office of Sustainability.

\subsection{Data Collection}
Beyond the recommendations, we engineered a full audio capture of the deliberations to create a high-quality dataset of transcripts for research, resulting in over 40 hours of audio. Each person wore a personal microphone for the duration of each assembly day. The microphones were remotely controlled by engineering staff to ensure synchronization. After the assembly, these audio files were aligned, merged, and cut. Then, they were transcribed into text using an AI speech-to-text tool and then diarized by speaker to create a dataset of 34 conversation transcripts spanning 6,300 speaker turns.

This study was approved by our university's Institutional Review Board (IRB). According to our IRB protocol, we are not able to publicly release the entire transcript dataset to protect delegates' privacy. We have received explicit consent from each delegate for all (anonymized) quotes included in this paper.

\section{From Ideas to Recommendations}
Using the above deliberation transcripts, our goal in this section is to empirically investigate: How might we trace the evolution and distillation of ideas into concrete recommendations within deliberative assemblies?
\begin{figure*}[t]
    \centering
    \includegraphics[width=0.75\linewidth]{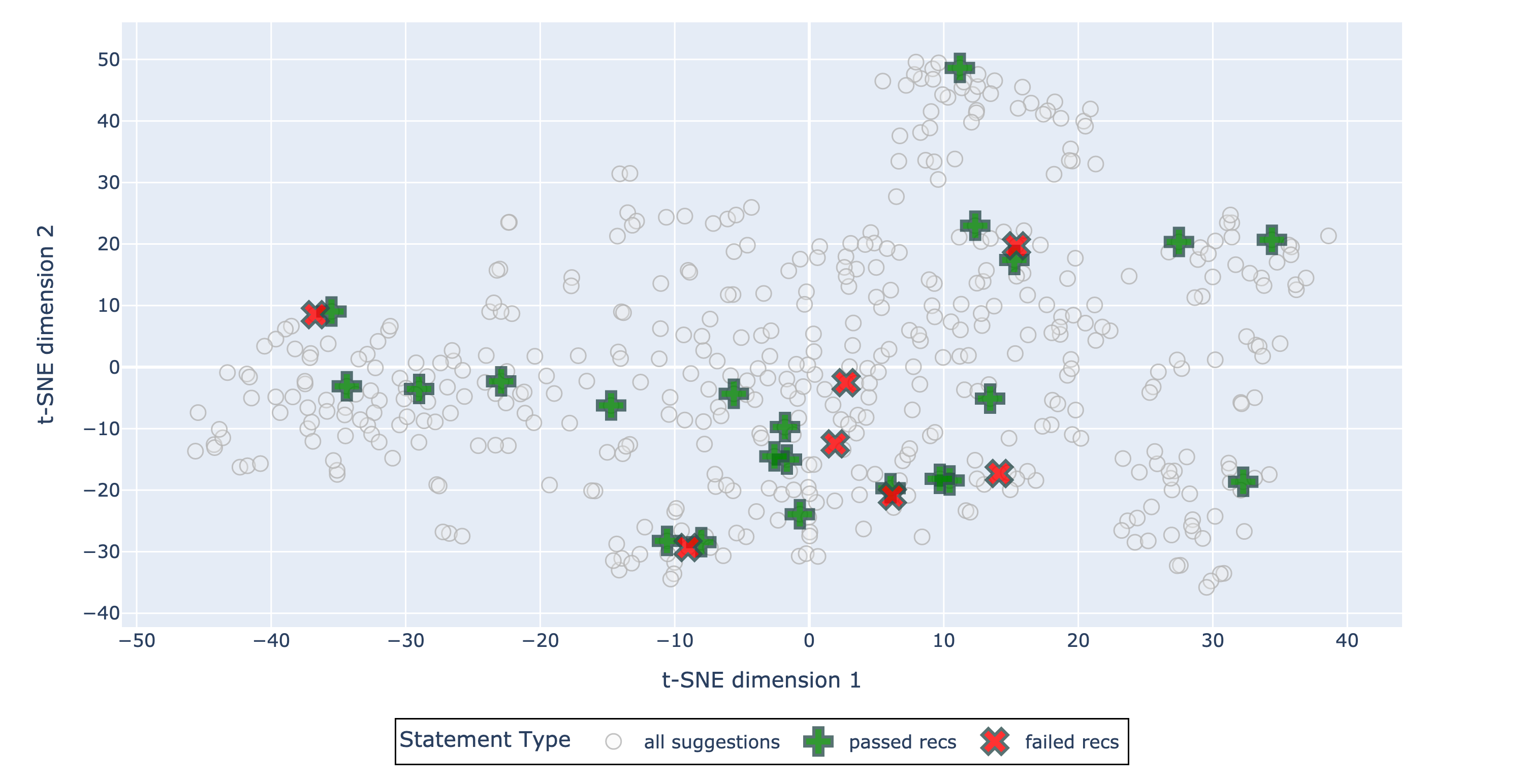}
    \caption{The embedding space of the extracted assembly suggestions and the final recommendations visualized in 2 dimensions using t-SNE. The gray circles represent the suggestions, the red X's represent final recommendations that ultimately failed in the final vote, and the green +'s represent final recommendations that passed the 80\% approval in the final vote.}
    \label{fig:suggestions-space}
\end{figure*}

\subsection{Identifying and visualizing suggestions}\label{sec:identifying}

We leverage LLMs to empirically trace and visualize how ideas---operationalized as explicit, voiced suggestions---evolve into final written recommendations during the assembly. First, we identify the total space of suggestions raised over the course of the assembly transcripts using \texttt{o3-mini} \citep{openai_openai_2025}. We iteratively refined the prompts in three rounds, manually evaluating on a subset of transcripts each time, resulting in 488 suggestions with nearly 0\% false negatives and a reduction of 90\% in false positives. Further evaluation details, exact prompts, and examples can be found in Appendix~\ref{appendix:promptengineering} and \ref{appendix:suggestion-prompts}.

Then we embed the suggestions using OpenAI's \texttt{text-embedding-3-small} and apply t-SNE to visually map suggestions and recommendations within a 2D semantic landscape (\Cref{fig:suggestions-space}). This visualization reveals notable \textit{gaps}—regions populated with suggestions that were not reflected in the final recommendations. While it is to be expected that the suggestions are a superset of the recommendations, these gaps warrant deeper investigation. 
We further investigate two hypotheses:
\begin{enumerate}[label=H\arabic*:,ref=H\arabic*,labelwidth=*,align=right]
    \item \label{h1} The gaps contain viable candidates for recommendations that were ultimately missed in the drafting process.
    \item \label{h2} The gaps contain suggestions pertaining to aspects of the assembly topic important enough for discussion, but during deliberation, these suggestions could not be transformed into viable recommendations. 
\end{enumerate}

In the case of \ref{h1}, this could be indicative of some shortcomings of the deliberative process during the assembly, and for which our framework could be used to create a tool or intervention in future tech-enhanced assemblies to prevent viable ideas from being missed and ensure the recommendations are comprehensive of all aspects of the topic. In the case of \ref{h2}, this might suggest that there are deliberative mechanisms influencing the collective decision to not draft recommendations on those aspects of the topic. Insights from these suggestions could then be leveraged to create a shareable assembly output to complement the final recommendations to better inform policymakers and other stakeholders, lending to increasing transparency and trust in the process. 

Of course, in reality, these two possibilities are not mutually exclusive, and our expectation is that we might find a combination of both. 

\subsection{Analyzing the missing suggestions.}\label{sec:missing}
\begin{figure*}
\centering
\includegraphics[width=0.6\linewidth]{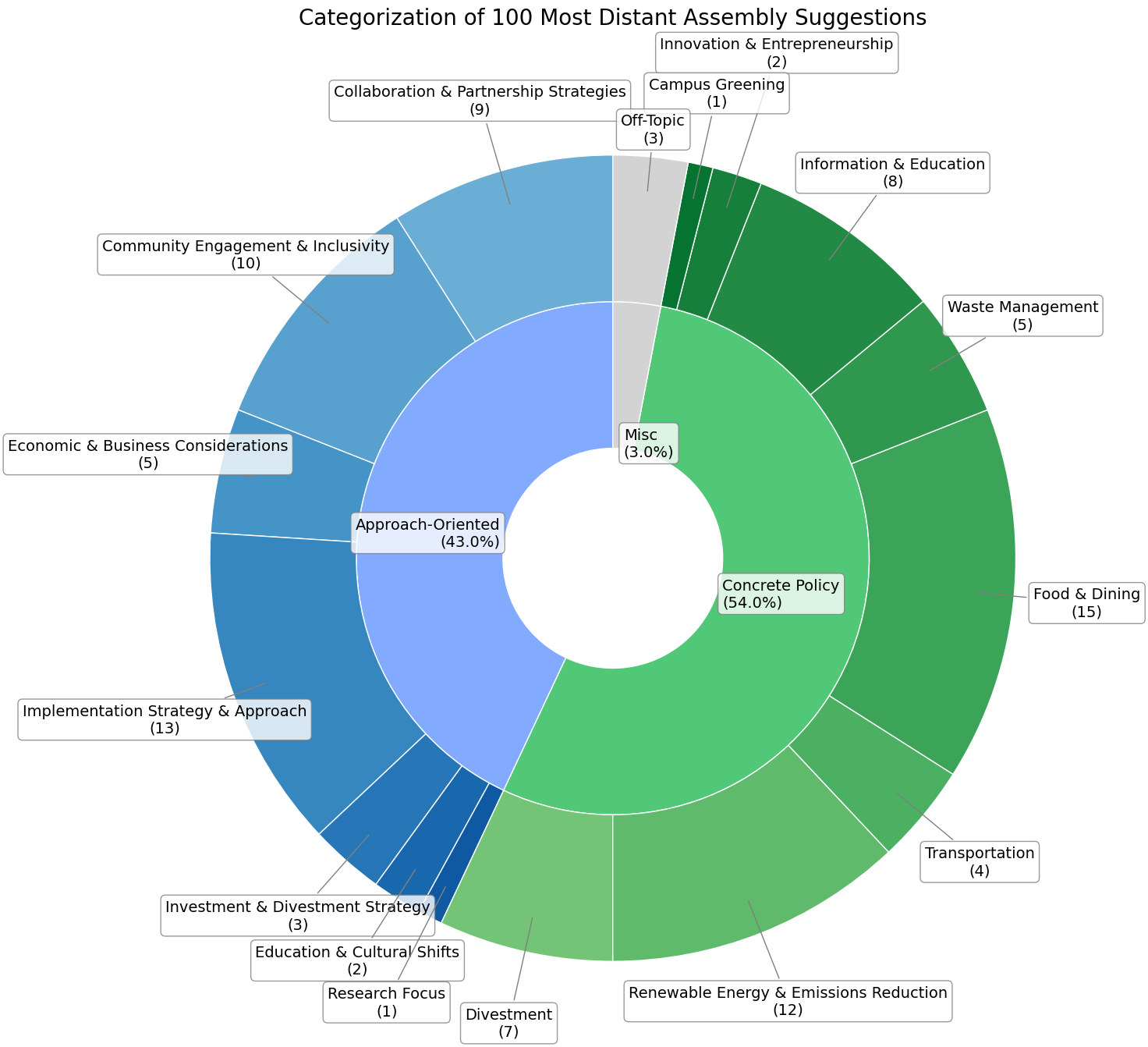}
\caption{Of the 100 most distant suggestions in the student assembly, 54 were concrete ideas and 43 focused on the approach to decision-making or implementation. The most common topics for the concrete suggestions were Food \& Dining, Renewable Energy, and Information \& Education.}
\label{fig:suggestions_categories}
\end{figure*}

\begin{figure*}
    \centering
    \includegraphics[width=0.75\linewidth]{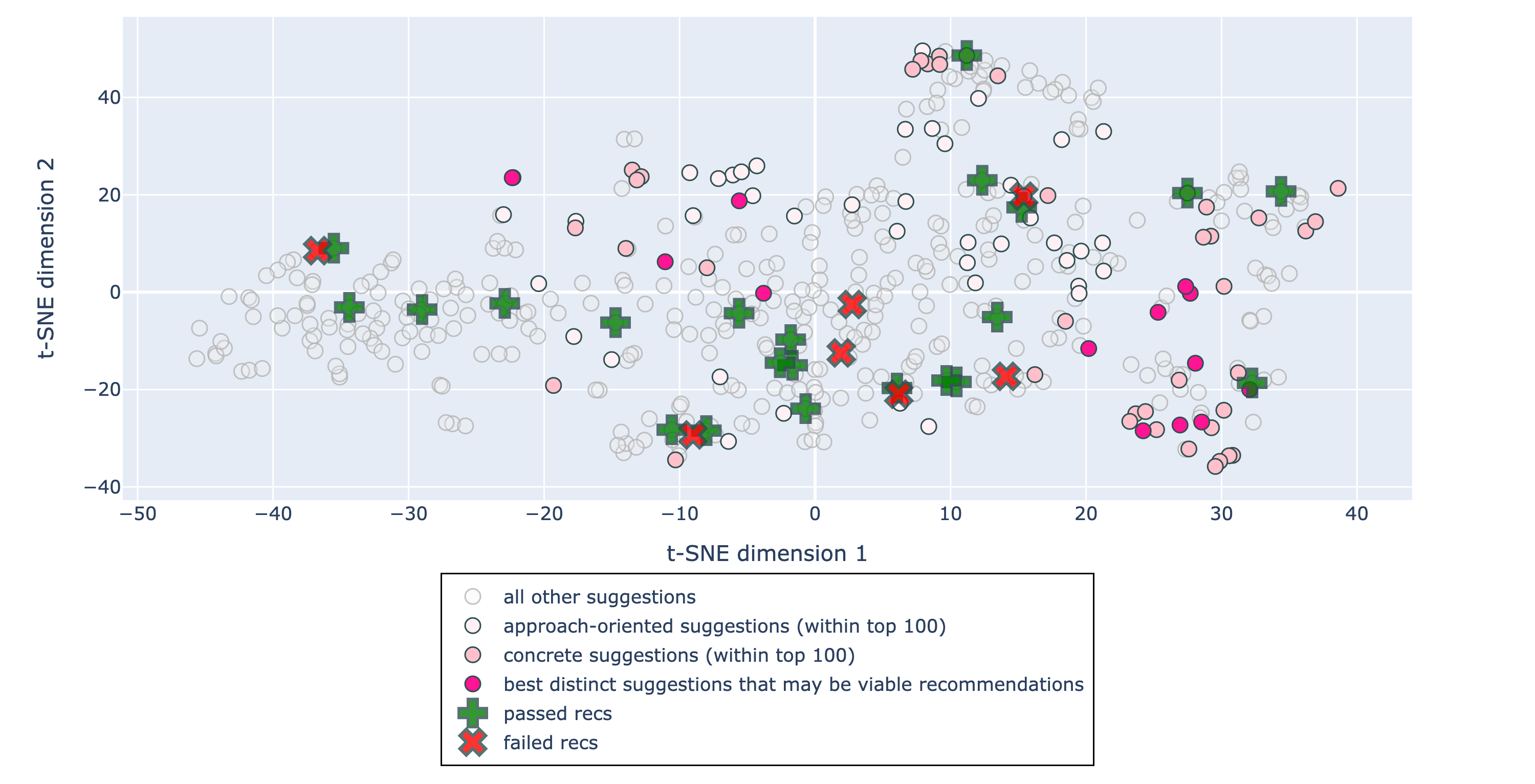}
    \caption{The same embedding space of the assembly suggestions and final recommendations visualized in 2 dimensions using t-SNE as in \Cref{fig:suggestions-space}, with the addition of the categorization of the suggestions in the gaps, i.e. the 100 suggestions most semantically distant from any final recommendation. Of these 100 suggestions, those that are approach-oriented and not concrete are colored white, the concrete suggestions are light pink, and the 13 concrete potentially viable suggestions are dark pink. We observe that the spatial gaps between the recommendations are mostly covered by these 100 suggestions. It is interesting to note that some areas have higher concentrations of more concrete suggestions (e.g., far right-hand side of the graph) whereas some regions have primarily approach-oriented suggestions (e.g., in the center of the graph).}
    \label{fig:suggestions-landscape}
\end{figure*}

We identify the 100 suggestions most semantically distant from final recommendations using cosine similarity between their embeddings. Qualitative analysis categorizes 54 as concrete policy suggestions (distinguishing specific actions, e.g., ``add solar panels,'' from general strategies, e.g., ``prioritize short-term goals'') and is visualized in Figure~\ref{fig:suggestions_categories}. From these, we manually\footnote{The decision to do this manually was to balance evaluation accuracy and time. Further motivation and discussion of this can be found in Appendix~\ref{appendix:manual}.} identified 13 distinct, actionable ideas that were overlooked in the final recommendations (\ref{h1}). The 13 suggestions were selected through manual qualitative review, ensuring they were both actionable and clearly distinct from existing recommendations, thus addressing genuine gaps overlooked during live deliberations. The detailed methodology with examples as well as the final list of 13 suggestions are in Appendix~\ref{appendix:student-suggestions}. 
The 46 remaining suggestions either overlap significantly with existing recommendations or convey delegate priorities and values rather than actionable policies (\ref{h2}), reflecting effective deliberation rather than genuine omissions (\Cref{fig:suggestions-landscape}).

These results demonstrate that while there are seemingly large gaps in the suggestion space, the vast majority of these are traces of effective and thorough deliberation in which many diverse ideas in various stages of concreteness can be generated and heard.

\subsection{Discussion}
Our findings highlight two key implications for future deliberative assemblies. First, identifying overlooked actionable suggestions (\ref{h1}) reveals areas where current deliberation practices can be improved. Our methodology could serve as the basis for live tools in future assemblies to ensure comprehensive consideration of all viable policy options. Second, recognizing that many suggestions reflect delegate values and priorities rather than concrete policies (\ref{h2}) underscores the importance of systematically capturing and communicating these insights to decision-makers. Sharing these perspectives alongside final recommendations can improve transparency and strengthen public trust in deliberative outcomes. 

Beyond tool development, applying our method across assemblies can surface broader patterns in idea dynamics and inform both theoretical and practical advancements in deliberative democracy.

\section{Detecting Delegate Stances over Time}
In this section, we describe the framework we designed to answer the second research question: How does the deliberative process shape delegate perspectives and influence voting dynamics over the course of the assembly? 

\begin{itemize}
\item {Delegate profiles}: We develop an LLM‐based framework for reconstructing each delegate’s evolving perspective across the stages of deliberation via compact, modular representations empirically grounded in the transcripts (\ref{sec:delegateprofiles}). 
    
\item Leveraging the profiles, we uncover \emph{when} and \emph{why} delegates revised their views, anchoring the drivers of vote change to the exact moments they emerged in the deliberative process (\ref{sec:profileanalysis}).
    
\item This framework enables synthesized insights to enhance assembly outputs and increase transparency (\ref{sec:implications}).
\end{itemize}

\subsection{Delegate Profile Design}\label{sec:delegateprofiles}
It is a difficult task to accurately understand or model the preferences and opinions of individuals from their speech alone across dozens of transcripts over the course of a deliberative assembly. 
There is an opportunity to leverage the rapidly improving capabilities of LLMs to synthesize the large-scale transcript data in a way that is useful for this task, but it remains paramount to do so in a way that (1) is maximally faithful to the delegates, (2) is systematically evaluable, and (3) minimizes the introduction of model biases. Moreover, in order to model change over time, there is the additional requirement of (4) modularity.

\subsubsection{Method}
To address these requirements, we propose a systematic framework for extracting each delegate's expressed opinions from the transcripts into a modular representation designed to be strongly grounded in the delegate's utterances, easy to evaluate or modify, and useful for a range of downstream analysis tasks.

The underlying intuition for constructing the delegate profiles is captured by: ``\textit{where in the transcript is there evidence for an individual’s expressed stance on a given statement?}" Our approach is: for an individual delegate and a 
statement, identify all instances in the transcripts in which that delegate speaks directly about—or responds to someone speaking directly about—the statement in a manner that reveals their stance (or lack thereof).
Repeating this for a collection of statements that cover important aspects of the central assembly topic results in a collection of evidence constituting a delegate's opinions. We call this a \textbf{delegate profile}.

Details regarding the prompts used in this section, including the various iterations on the prompt and evaluations thereof, are included in Appendices \ref{appendix:profileprompts}, \ref{appendix:delprofeval}, and \ref{appendix:delegateprofileprompts}.

\subsection{Tracing Stance Change via Delegate Profiles}\label{sec:profileanalysis}

Here, we illustrate how these profiles let us track exactly \emph{when} and \emph{why} delegates change their minds. We look at the case of a delegate profile showcasing their opinion change on a particular assembly recommendation. The drafted version on Day 2 was  ``Divest the endowment from fossil fuels.'' This was revised on Day 3 to ``Divest the endowment from fossil fuels within the next 5 years.'' This delegate initially voted Agree on Day 2, but changed to Neutral\footnote{In voting, delegates could vote Agree, Neutral, or Disagree. Support for recommendations-in accordance with previous assemblies-was calculated by \# Agrees / total. Therefore, changing from Agree to Neutral has the same effect as changing to Disagree.} on Day 3 (when expert input reveals feasibility concerns).

In each box below are the speaker turns identified as relevant by \texttt{o3-mini} in chronological order. For brevity, we have omitted the full LLM output and italicized the most relevant parts of each quote.
\vspace{1em}
\begin{mdframed}[frametitle={Assembly Day 1},
    frametitlefont=\bfseries,
    innertopmargin=1em
]
Discuss Pre-Reads Breakout:
\begin{itemize}
    \item ``...it would not be worth divesting immediately if it's something that like takes... funding away from like climate research or like supporting lower income students, …So I think that it definitely wouldn't be worth it to do it like very abruptly in the short term. But \textit{I think that it could also be done in a way like over a longer time frame}..."
    \item ``Like, basically, if it's truly no strings Attached, then \textit{it's definitely worth like it's just going to be a net positive [f]or climate research as a whole}. But it's very rare that like it is truly no strings attached." 
    \item  ...And then we talked a little bit about the value of like divesting from fossil fuels and like how we kind of felt about that. And I think we kind of said that \textit{it wasn't really worth immediately divesting from fossil fuels because it would have a lot of negative impacts on like funding} for programs like supporting lower income students and like supporting like just various different research, like especially climate research. And that \textit{it would be like a better idea to divest more slowly over time in a way that isn't going to impact funding for those things}. Then we talked a little bit about like some of the universities that have been able to successfully divest from fossil fuels. 
\end{itemize}

\noindent Brainstorm Questions for Experts Breakout:
\begin{itemize}
    \item ``...Like, like \textit{trying to divest more from fossil fuel companies. Like how feasible is that?} Like it seems like it should be feasible over like longer time frame. We can kind of get that, get a better sense of that from the experts."
\end{itemize}
\end{mdframed}

\begin{mdframed}[frametitle={Assembly Day 2},
    frametitlefont=\bfseries,
    innertopmargin=1em
]
Draft Initial Recommendations Breakout:

\begin{itemize}
    \item “…I was kind of curious about the like since like the fossil fuels \textit{seems to be like such a small portion of the endowment}.”
    \item “...I feel like \textit{even like if it doesn't make that much of a difference to the endowment, it'd be great just as like like building awareness type of thing}.”
    \begin{itemize}
        \item The above was a response to another delegate: ``Like if we could just like completely divest."
    \end{itemize}
\end{itemize}
	
\textbf{Day 2 Voting: Agree}
\end{mdframed}

\begin{mdframed}[frametitle={Assembly Day 3},
    frametitlefont=\bfseries,
    innertopmargin=1em
]

Info Fair with [Expert]
\begin{itemize}
    \item I think I remember mentioned during the panel that like, \textit{only a very small percentage of the endowment was actually related to fossil fuels}. So I was curious after that about like, \textit{would it be feasible} then to actually like, fully divest the endowment from fossil fuels? \textit{To make more of a statement, honestly, than anything else, just to kind of show that MIT is like, actually leading by example} and like, if so, how long would it take? And also, like, \textit{would it have a significant impact on the endowment?}
\end{itemize}

\noindent Info Fair Debrief Breakout
\begin{itemize}
    \item ``Yeah. Actually like, we talked about divesting from fossil fuels a lot too, and [Expert] kind of made similar points about like, since \textit{it is such a minuscule portion of the endowment, it's really not. It's not going to impact} those companies negatively if we were to divest from them. And it's also going to be \textit{generally just kind of difficult} to do and \textit{not very necessary} either because it is such a small portion. \textit{So that also kind of flipped my perspective on divesting}."
    \item ``Yeah, I think I would definitely tweak a couple of my recommendations to be more focused on the communication aspect of things, but also probably think what my. Recommendations was to Divest from. Or divest the endowment from fossil fuel. And \textit{I would probably either change or. Just remove that recommendations. And I think after talking with [Expert], it seems like less of a priority}."
    \item ``I think... \textit{the divesting from fossil fuels, the lower down of the priority list} for us."
\end{itemize}
	
\textbf{Day 3 Voting: Neutral}
\end{mdframed}
\vspace{1em}
\par On Days 1 and 2, this delegate consistently expressed support for divestment noting that ``it would be like a better idea to divest more slowly over time in a way that isn't going to impact funding for programs like supporting lower income students and... climate research'' and that ``if it’s truly no strings attached, then it’s definitely worth it.'' They mentioned that ``even like if it doesn't make much of a difference,'' it is worth pursuing ``to make more of a statement'' and to ``show that MIT is like, actually leading by example,'' thus voting \emph{Agree} on Day 2.  However, during the Day 3 Info Fair debrief, a conversation with an expert revealed that fossil fuels represented ``such a minuscule portion of the endowment,'' making divestment ``generally just kind of difficult'' and ``not very necessary.'' This new feasibility insight prompted this delegate to flip their vote to \emph{Neutral}, and in the post‐survey they selected ``My opinion changed on this topic." 
This delegate was not alone in removing their support for this recommendation, which passed on Day 2, causing it to fall below the 80\% threshold on Day 3. Six other delegates changed from Agree to Neutral/Disagree, three also citing opinion change and three indicating that the central question affected their voting strategy. 
This case illustrates how direct expert input can trigger a clear pivot in delegate opinions when new practical constraints emerge.

\subsection{Discussion}\label{sec:implications}
Deliberative assembly output is traditionally communicated by sharing the final passed recommendations to decision-makers. However, this is fundamentally limited and hides the nuance and depth of the deliberative process.
Our framework presents a promising amendment for this issue by partially automating the process of synthesizing the delegates' underlying values,  priorities, and recommendation transformations, thus reducing the complexity of such a task and enabling increased transparency into critical insights that are otherwise invisible to the authorities receiving the recommendations.

For example, the recommendation regarding MIT's divestment from fossil fuels passed on Day 2 but fell short of 80\% approval on Day 3 and was not passed on to decision-makers post-assembly.
This risks falsely implying that the delegates overall did not support divestment. In reality, as illustrated by the delegate profile, expert input revealed practical constraints and changed their opinion on feasibility, but the delegate strongly valued divestment in principle. This delegate was not alone in withdrawing support for this recommendation for the same reasons; six other delegates retracted their agree votes on Day 3. The profile highlights the nuanced trade-off between delegates wanting to make a value-driven statement in recommending divestment versus prioritizing only feasible recommendations which would otherwise remain opaque to the decision-makers.

\section{Limitations \& Future Directions}\label{sec:idealimitations}
As mentioned before, due to time constraints, it was not feasible in the scope of this work to analyze the full set of 488 suggestions and instead we focused our analysis on a subset of the 100 most distant suggestions. By nature of selecting the most distant ones, we maximized the likelihood of finding those that are distinct from the recommendations.  
It would be worthwhile to test to what extent calculating distance in the 2D t-SNE space (or other spaces resulting from alternative dimensionality reduction techniques) impacts the ordered list of distant suggestions and visualized clusters.

Within the scope of this work, it was not possible to recover the true reasons why these specific suggestions were missed. However, the methodology and prompting framework we developed could be leveraged in future work to design a tool for live assemblies. Such a tool could help capture these promising ideas in real-time and remind the delegates of them during the drafting phase, thereby improving the deliberative assembly process and outcomes.

Future work could also leverage insights from the discarded suggestions to create a shareable assembly output to complement the final recommendations to better inform policymakers and other stakeholders, increasing transparency and trust in the assembly process. 

Beyond tool development, a natural next step is to apply this framework to other deliberative assemblies or public forums. 
Comparative applications could investigate whether other assemblies exhibit similar degrees of coverage in the suggestion space, or whether missed ideas cluster differently depending on topic, group size, or facilitation style. These extensions would allow the framework to serve not only as a support tool for future assemblies, but also as a foundation for richer cross-assembly analysis and research on deliberative effectiveness. 

\section{Conclusion}

In this work, we leverage recent advances in LLMs to develop empirical methods that trace idea evolution in deliberative assemblies, reconstruct delegate stances, and explain how deliberation shapes which proposals ultimately succeed.

First, we asked: How might we trace the evolution and distillation of ideas into concrete recommendations within deliberative assemblies? In response, we developed a framework for identifying and analyzing the full space of suggestions voiced during a deliberative assembly. Using LLMs to extract, embed, and visualize suggestions from transcript data, we revealed large gaps between initial ideas and final recommendations. Yet contrary to appearances, these gaps often signaled healthy deliberative processes—where ideas were surfaced, refined, or appropriately discarded. Still, a subset of missed-but-viable ideas showed the limits of current practices and pointed toward promising intervention tools.

Second, we asked: How does the deliberative process shape delegate perspectives and influence voting dynamics over the course of the assembly? In response, we introduced a systematic framework for reconstructing delegate stances from the transcripts. This method anchors vote changes to the precise moments and deliberative mechanisms that produced them. We also demonstrated how these profiles can enhance the transparency of assembly outcomes by surfacing the reasoning behind votes—critical insights otherwise hidden from decision-makers.

Our framework lays the groundwork for future deliberative assemblies to develop real-time support tools that capture overlooked suggestions and systematically surface the underlying rationale behind policy recommendations. In doing so, it lays the groundwork for future human-centered tools that can capture, synthesize, and share evolving stances—without losing the richness and nuance of the deliberative process.

\bibliography{references}

\newpage
\appendix
\setcounter{secnumdepth}{2} 

\section*{Appendix Overview}
\begin{itemize}
    \item \Cref{appendix:promptengineering} contains the details on prompt iteration and evaluation for the suggestions.
    \item \Cref{appendix:suggestion-prompts} contains the actual prompts and Pydantic schemas for the suggestions.
    \item \Cref{appendix:student-suggestions} contains the full final list of the 13 missing yet viable suggestions discovered in the paper.
    \item \Cref{appendix:profileprompts} and \Cref{appendix:delprofeval} contain the prompt iteration and evaluation details (respectively) for the delegate profiles.
    \item \Cref{appendix:delegateprofileprompts} contains the actual prompts and Pydantic schemas for the delegate profiles.
\end{itemize}

\section{Suggestions Prompt Iteration \& Evaluation Details}\label{appendix:promptengineering}
Through a systematic iterative approach to prompt engineering, we significantly improved the extraction of actionable sustainability suggestions. In the first stages of prompt engineering, we focused on the breakout sessions in which the delegates drafted recommendations on Day 2 because these conversations explicitly contained an abundance of suggestions for recommendations. To evaluate, we read through the conversations entirely and manually inspected model outputs for each speaker turn, noting any instances of false positives or negatives.

The initial prompt (\Cref{appendix:suggpromptv0}) often produced false positives, such as questions misinterpreted as proposals (e.g., ``Is there an equivalent for [university program]?"), factual statements interpreted normatively (e.g., ``There's no Department of Climate"), or procedural comments (e.g., ``Let's screenshot that"). we did not find any clear cases in which the model missed a suggestion (i.e. a false negative).

To combat the false positives while preserving true suggestions, we added filtering instructions for questions, facts, and procedural mentions in the second prompt version (\Cref{appendix:suggpromptv1}). This significantly improved the outputs, however the model still struggled slightly with  distinguishing procedural from substantive content. Moreover, as opposed to other text data, conversation transcripts tend to contain more ambiguous statements, sentences that are cut off, or split between several speaker turns due to interruptions or trailing off. we observed that the LLM classifications of such utterances was overly liberal, so we refined the prompt to include more explicit instructions that emphasized completeness, specificity, and leveraging conversation context when determining if a statement is truly a suggestion. Lastly, we added examples of the procedural statements and other types false positives we saw frequently in the final prompt version (\Cref{appendix:suggpromptv2}).

In order to evaluate a larger set of the suggestions across a more diverse set of conversations (one Day 2 drafting breakout, one Day 1 pre-read debrief breakout, and one Day 3 info fair debrief breakout), we leveraged Anthropic's Claude 3.7 Sonnet\footnote{This model was chosen as an evaluator because it is a more powerful state-of-the-art LLM more suitable for this type of evaluation task (but too expensive for doing the entire suggestion extraction). To avoid any potential self-preference bias \cite{panickssery_llm_2024}, it was also important to choose an LLM from a different family of models.} to analyze the outputted suggestions of \texttt{o3-mini} before and after the second prompt refinement. One conversation at a time, we tasked Claude with estimating the number of false positives and to give an overall analysis of the remaining errors:
\begin{itemize}
    \item ~90\% reduction in false positives
    \item Higher recall of genuine sustainability suggestions
    \item Strong focus on institutional actions rather than group process
    \item Improved handling of complex, nuanced suggestions
\end{itemize}
We manually inspected Claude's output as a sanity check and found that the analysis was thorough and accurate. The full text of all prompt versions used to extract the suggestions for the entire assembly can be found in \Cref{appendix:suggestion-prompts} and the Claude analysis output comparing prompt versions can be found in \Cref{appendix:claude-analysis}.

\subsection{Suggestions Prompt Evaluation with Claude}\label{appendix:claude-analysis}
In all prompts below, all delegate names have been redacted for anonymity. 

Comparing the initial prompt (\ref{appendix:suggpromptv0}) and the second iteration (\ref{appendix:suggpromptv1}):

\begin{mdframed}
\# Error Analysis of Revised Output

\#\# Overview

The revised prompt has significantly improved the quality of the output, with far fewer false positives and a more accurate representation of genuine suggestions. Let me analyze the results:

\#\# True Positives

The model still correctly identifies most legitimate suggestions, such as:

- Delegate 10's suggestion about avoiding creating a new organization

- Delegate 19's suggestions about centralizing student organizations

- Delegate 4's proposal about divesting and establishing a strong central program

- Delegate 13's suggestion about contacting heads of student organizations

- Delegate 7's ideas about increasing sustainability awareness

\#\# False Positives Analysis

The revised output has dramatically reduced false positives, but some minor issues remain:

1. **Vague or ambiguous statements still occasionally treated as suggestions**:

- Delegate 19: "Or the other groups that we should include." → This could be a question or observation rather than a clear suggestion.

- Delegate 4: "It'd be cool if we have something bigger than that." → This is more of a wish than a concrete suggestion.

- Delegate 4: "Maybe we should write that down."

→ This is procedural rather than a sustainability suggestion.

- Delegate 4: "You can use my phone." → This is procedural and not related to the assembly goal.

2. **Context-dependent statements**:

- Delegate 4: "Go ask the philosophers. Imagine coming to MIT and being like." → The context suggests this might be somewhat sarcastic rather than a genuine suggestion.

\#\# False Negatives

Without seeing the full transcript, it's hard to identify all potential false negatives. However, the model appears to be more conservative now, which might mean some subtle suggestions could be missed if they're not explicitly phrased as recommendations.

\#\# Improvements from Previous Version

1. **Questions properly filtered out**:

- Questions like "Is there an equivalent for [university program], but, like, local?" are no longer misinterpreted as suggestions.

2. **Statements of fact excluded**:

- Statements like "Yeah, but like there's no Department of Climate" are no longer transformed into suggestions.

3. **Simple identifications properly filtered**:

- Entity mentions like "The Environmental Solutions Initiative" and "MCSC" are no longer treated as suggestions.

4. **Duplicates reduced**:

- The repeated suggestions from the same quotes have been eliminated.

5. **More faithful paraphrasing**:

- The paraphrases stick closer to the original meaning without adding interpretations.

\#\# Remaining Issues

1. **Procedural statements**: Some procedural suggestions that don't directly relate to the sustainability goal are still being included (e.g., "I suggest using my phone").

2. **Informal language**: The model still sometimes struggles with casual or ambiguous phrasing.

3. **Incomplete suggestions**: Some suggestions that are cut off mid-sentence are still included, though the model is more careful about not extrapolating beyond what was said.

\#\# Overall Assessment

The revised prompt has led to a significantly improved output with:

1. Far fewer false positives

2. More faithful paraphrasing

3. Better discrimination between suggestions and other speech acts

4. Elimination of duplicate processing

The remaining issues are relatively minor and likely cannot be completely eliminated without risking an increase in false negatives.
\end{mdframed}

Comparing the second prompt (\ref{appendix:suggpromptv1}) with the final prompt (\ref{appendix:suggpromptv2}):

\begin{mdframed}
\# Error Analysis of Final Prompt Result

\#\# Overview
The updated prompt has successfully filtered out most procedural language and ambiguous suggestions, resulting in a much cleaner and more focused set of relevant sustainability suggestions. This version shows significant improvement over both previous iterations.

\#\# True Positives
The model continues to correctly identify legitimate sustainability suggestions, such as:
- Delegate 10's suggestion about an accessible disclosure system for MIT's funding
- Delegate 19's proposals about centralizing sustainability efforts under one office
- Delegate 4's suggestions about tracking sustainability goals
- Delegate 13's idea about a sustainability competition modeled after motorsports
- Delegate 7's suggestion about creating a centralized list for sustainability opportunities

\#\# False Positives Analysis
The revised output has very few false positives remaining, but there are still some minor issues:

1. **Still some vague or potentially procedural suggestions**:
   - Delegate 4: "Should we just recommend both of them? We could, we could just put for both of them. They can work together." → This appears to be about the process of making recommendations rather than a concrete sustainability suggestion for MIT.
   - Delegate 13: "I guess somebody confused like. So we're advocating for adding an additional class." → This is somewhat ambiguous and lacks specificity about what kind of class.

2. **Potentially incomplete suggestions**:
   - Delegate 4: "I propose that MIT should change sustainability for engineers." → This is vague about what exactly should be changed.
   - Delegate 10: "I propose that MIT should include sustainability-related courses in the restricted elective list." → This is from a partial quote "Related courses into the restricted elective list" which might lack context.

\#\# False Negatives
It appears the prompt may be appropriately filtering out procedural, ambiguous, and incomplete suggestions that don't constitute clear policy recommendations. There's no obvious indication of legitimate sustainability suggestions being missed.

\#\# Improvements from Previous Versions

1. **Procedural language effectively filtered**:
   - Previous examples like "let's write this down," "you can use my phone," and "let's screenshot that" are now correctly excluded.

2. **Focus on MIT policies vs. group discussion**:
   - The output now focuses on concrete actions MIT could take rather than how the group should proceed.

3. **Greater specificity in suggestions**:
   - Most suggestions are now specific and actionable rather than vague.

4. **Better context retention**:
   - The model is better at including sufficient context in longer quotes to ensure the suggestion makes sense.

5. **Reduced redundancy**:
   - Fewer duplicative or overlapping suggestions appear in the output.

\#\# Remaining Minor Issues

1. **A few suggestions about recommendation process**:
   - There are still a couple of instances where the line between procedural discussion and sustainability suggestion is blurry.

2. **Some suggestions could be more specific**:
   - A few suggestions remain somewhat vague about implementation details.

\#\# Overall Assessment

The revised prompt has resulted in an excellent output with:
1. Very few false positives
2. Good coverage of genuine sustainability suggestions
3. Clear distinction between discussion process and policy recommendations
4. Appropriate filtering of ambiguous or incomplete suggestions
\end{mdframed}

\section{Suggestions Prompts}\label{appendix:suggestion-prompts}

\subsection{Initial Prompts}\label{appendix:suggpromptv0}

\begin{mdframed}[frametitle={System Instructions:},
    frametitlefont=\bfseries,
    innertopmargin=1em
]
\begin{lstlisting}
You are analyzing transcripts from a deliberative assembly. Your task is to identify suggestions or ideas raised in relation to the central assembly goal.

You will be given MULTIPLE UTTERANCES to evaluate at once. For EACH utterance, determine if it contains a suggestion or idea related to the goal: "What policies or initiatives should MIT prioritize to enhance campus sustainability and meet its climate commitments? How can we, as students, use our experiences to inform sustainability action at MIT? What actions, partnerships, or projects align with student values and also could have the biggest impact across the campus? The assembly aims to generate actionable recommendations that position MIT as a leader in campus sustainability and serve as a model for other institutions."

For each utterance:
1. If it contains a relevant suggestion/idea:
   - Record the *exact* quote verbatim containing it
   - Write a faithful but concise paraphrase of the suggestion so that it can be understood standalone in the first person (eg "I think X" or "Y should be implemented because", etc)
2. If it does NOT contain a relevant suggestion/idea:
   - Leave both the quote and suggestion fields empty

You must evaluate EACH utterance independently and maintain the same order as provided in the input.
\end{lstlisting}
\end{mdframed}

\begin{mdframed}[frametitle={Prompt:},
    frametitlefont=\bfseries,
    innertopmargin=1em
]
\begin{lstlisting}
## **Assembly Goal:**
"What policies or initiatives should MIT prioritize to enhance campus sustainability and meet its climate commitments? How can we, as students, use our experiences to inform sustainability action at MIT? What actions, partnerships, or projects align with student values and also could have the biggest impact across the campus? The assembly aims to generate actionable recommendations that position MIT as a leader in campus sustainability and serve as a model for other institutions."

## **Current utterances to evaluate:**
{batch}

## **Instructions**
For EACH utterance in the list above:

1. You MUST return classifications in the SAME ORDER as the utterances above.

2. For each utterance, determine:
   - Whether it contains a relevant suggestion/idea
   - The *exact* quote verbatim containing it (if relevant)
   - A faithful but concise paraphrase of the suggestion so that it can be understood standalone in the first person, e.g. "I think X" or "Y should be implemented because", etc. (if relevant)
   - Leave fields empty if not relevant

3. Return a JSON list matching the specified schema, with one entry per utterance.
{schema_string}

CRITICAL REMINDERS:
- Your output list MUST include ALL input utterances in the same order
- Evaluate each utterance independently
- Only include suggestions specifically related to the assembly goal
- If a single utterance contains several unrelated ideas/suggestions, include a separate paraphrased sentence for each.
- If in doubt, leave fields empty
\end{lstlisting}
\end{mdframed}

\subsection{Second Iteration Prompts}\label{appendix:suggpromptv1}
\begin{mdframed}[frametitle={System Instructions:},
    frametitlefont=\bfseries,
    innertopmargin=1em
]
\begin{lstlisting}
You are analyzing transcripts from a deliberative assembly. Your task is to identify ONLY explicit suggestions or ideas raised in relation to the central assembly goal.

You will be given MULTIPLE UTTERANCES to evaluate at once. For EACH utterance, determine if it contains a clear, explicit suggestion or idea related to the goal: "What policies or initiatives should MIT prioritize to enhance campus sustainability and meet its climate commitments? How can we, as students, use our experiences to inform sustainability action at MIT? What actions, partnerships, or projects align with student values and also could have the biggest impact across the campus? The assembly aims to generate actionable recommendations that position MIT as a leader in campus sustainability and serve as a model for other institutions."

For each utterance:
1. If it contains a CLEAR, EXPLICIT suggestion/idea (not just a question, observation, or naming of an entity):
   - Record the *exact* quote verbatim containing it
   - Write a faithful but concise paraphrase of the suggestion so that it can be understood standalone in the first person (eg "I think X" or "Y should be implemented because", etc)
   - The paraphrase should NOT add any new ideas or interpretations not present in the original quote
   
2. If it does NOT contain a relevant suggestion/idea OR if it only contains:
   - Questions seeking information (e.g., "Is there X?")
   - Simple statements of fact without proposals (e.g., "There is no X.")
   - Mere identification or naming of entities without suggesting action
   - General observations without concrete proposals
   - Leave both the quote and suggestion fields empty

Critically, DO NOT interpret questions, statements of fact, or mentions of entities as suggestions unless they explicitly propose action.

You must evaluate EACH utterance independently and maintain the same order as provided in the input.
\end{lstlisting}
\end{mdframed}

\begin{mdframed}[frametitle={Prompt:},
    frametitlefont=\bfseries,
    innertopmargin=1em
]
\begin{lstlisting}
## **Assembly Goal:**
"What policies or initiatives should MIT prioritize to enhance campus sustainability and meet its climate commitments? How can we, as students, use our experiences to inform sustainability action at MIT? What actions, partnerships, or projects align with student values and also could have the biggest impact across the campus? The assembly aims to generate actionable recommendations that position MIT as a leader in campus sustainability and serve as a model for other institutions."

## **Current utterances to evaluate:**
{batch}

## **Instructions**
For EACH utterance in the list above:

1. You MUST return classifications in the SAME ORDER as the utterances above.

2. For each utterance, determine:
   - Whether it contains a CLEAR, EXPLICIT suggestion/idea (not just a question, observation, or naming of an entity)
   - The *exact* quote verbatim containing it (if relevant)
   - A faithful but concise paraphrase of the suggestion that DOES NOT add interpretations beyond what's in the original text

3. DO NOT classify as suggestions:
   - Questions seeking information (e.g., "Is there a program for X?")
   - Simple statements of fact (e.g., "There is no Office of X")
   - Mere identification or naming of entities (e.g., "The Environmental Solutions Initiative")
   - General observations without concrete proposals
   - Statements noting lack of decisions (e.g., "We haven't decided on X")
   
4. If a transcript contains duplicate utterances, ensure you analyze each instance independently

5. Return a JSON list matching the specified schema, with one entry per utterance.
{schema_string}

CRITICAL REMINDERS:
- Your output list MUST include ALL input utterances in the same order
- Evaluate each utterance independently
- Only include EXPLICIT suggestions specifically related to the assembly goal
- When paraphrasing, DO NOT transform descriptive statements into normative ones
- DO NOT interpret questions as suggestions
- DO NOT interpret mere mentions of entities or initiatives as suggestions for their implementation
- If a single utterance contains several unrelated ideas/suggestions, include a separate paraphrased sentence for each
- If in doubt, leave fields empty
\end{lstlisting}
\end{mdframed}

\subsection{Final Prompts}\label{appendix:suggpromptv2}
Below are the final prompts we used to extract the suggestions from the transcripts using OpenAI's \texttt{o3-mini} model:

\begin{mdframed}[frametitle={System Instructions:},
    frametitlefont=\bfseries,
    innertopmargin=1em
]
\begin{lstlisting}
You are analyzing transcripts from a deliberative assembly. Your task is to identify ONLY explicit suggestions or ideas raised in relation to the central assembly goal about MIT sustainability policies and initiatives.

You will be given MULTIPLE UTTERANCES to evaluate at once. For EACH utterance, determine if it contains a clear, explicit suggestion or idea related to the goal: "What policies or initiatives should MIT prioritize to enhance campus sustainability and meet its climate commitments? How can we, as students, use our experiences to inform sustainability action at MIT? What actions, partnerships, or projects align with student values and also could have the biggest impact across the campus? The assembly aims to generate actionable recommendations that position MIT as a leader in campus sustainability and serve as a model for other institutions."


For each utterance:
1. If it contains a CLEAR, EXPLICIT, COMPLETE suggestion/idea that:
   - Directly relates to MIT sustainability policies or initiatives (not just group process)
   - Is fully articulated (not cut off mid-thought)
   - Proposes a concrete action for MIT as an institution (not just for the discussion group)
   - Then record the *exact* quote verbatim containing it and write a faithful paraphrase
   
2. DO NOT include as a suggestion if the utterance:
   - Is procedural (e.g., "let's write this down," "we should screenshot this," "you can use my phone")
   - Is about the process of formulating recommendations (e.g., "we should ask someone," "let's make a list")
   - Is ambiguous or lacks sufficient context to determine the specific recommendation
   - Is incomplete or cut off mid-thought
   - Is a question seeking information without proposing action
   - Is a simple statement of fact without a proposal
   - Only identifies or names entities without suggesting action
   - Is a general observation without concrete proposals
   
The paraphrase should NOT add any new ideas or interpretations not present in the original quote. CopyThe paraphrase MUST be written in the first person (e.g., "I suggest that MIT should..." or "I propose that we implement..."), regardless of how the original statement was phrased. You must evaluate EACH utterance independently and maintain the same order as provided in the input.
\end{lstlisting}
\end{mdframed}

\begin{mdframed}[frametitle={Prompt:},
    frametitlefont=\bfseries,
    innertopmargin=1em
]
\begin{lstlisting}
## **Assembly Goal:**
"What policies or initiatives should MIT prioritize to enhance campus sustainability and meet its climate commitments? How can we, as students, use our experiences to inform sustainability action at MIT? What actions, partnerships, or projects align with student values and also could have the biggest impact across the campus? The assembly aims to generate actionable recommendations that position MIT as a leader in campus sustainability and serve as a model for other institutions."

## **Current utterances to evaluate:**
{batch}

## **Instructions**
For EACH utterance in the list above:

1. You MUST return classifications in the SAME ORDER as the utterances above.

2. For each utterance, determine:
   - Whether it contains a CLEAR, EXPLICIT, COMPLETE suggestion/idea specifically about MIT sustainability policies or initiatives
   - The *exact* quote verbatim containing it (if relevant)
   - A faithful but concise paraphrase of the suggestion that DOES NOT add interpretations beyond what's in the original text and MUST be written in the first person ("I suggest/propose/recommend that...")

3. DO NOT classify the following as suggestions:
   - Procedural statements about the discussion itself (e.g., "let's write this down," "we should make a list")
   - Comments about how to gather information (e.g., "we should ask experts," "let's compile data")
   - Discussion of group process (e.g., "let's split up tasks," "we should vote on this")
   - Questions seeking information (e.g., "Is there a program for X?")
   - Simple statements of fact (e.g., "There is no Office of X")
   - Mere identification or naming of entities (e.g., "The Environmental Solutions Initiative")
   - General observations without concrete proposals
   - Statements noting lack of decisions (e.g., "We haven't decided on X")
   - Incomplete thoughts or statements cut off mid-sentence
   - Ambiguous statements without clear meaning in isolation
   - Suggestions about the assembly process rather than about MIT sustainability policies

4. ONLY include suggestions that:
   - Propose concrete actions towards and directly related to the Assembly Goal
   - Are complete and clearly articulated
   - Would make sense as a standalone recommendation outside the current conversation

5. Return a JSON list matching the specified schema, with one entry per utterance.
{schema_string}

CRITICAL REMINDERS:
- Your output list MUST include ALL input utterances in the same order
- Evaluate each utterance independently
- Only include EXPLICIT suggestions specifically related to MIT sustainability policies or initiatives
- When paraphrasing, DO NOT transform descriptive statements into normative ones
- DO NOT include procedural suggestions about the discussion process itself
- DO NOT include suggestions about information gathering unless they propose a specific MIT policy
- DO NOT include incomplete or ambiguous statements
- If a single utterance contains several unrelated ideas/suggestions, include a separate paraphrased sentence for each
- ALWAYS paraphrase suggestions in the first person (e.g., "I suggest that MIT should...")
- If in doubt, leave fields empty
\end{lstlisting}
\end{mdframed}

For all prompt versions, the \texttt{batch} contains 10 speaker turns from a transcript, and the \texttt{schema\_string} specifies the output formatting, which is a list of JSONs each containing the original speaker name and utterance from the transcript as well as a paraphrasing of the suggestion (which is left empty if no the utterance does not contain any suggestions). The schema is defined by the Pydantic model below with a field validator to ensure that the model returns a classification for all 10 speaker turns that were input.

\begin{mdframed}[frametitle={Model Schema:},
    frametitlefont=\bfseries,
    innertopmargin=1em
]
\begin{lstlisting}[language=Python,showstringspaces=false]
class SuggestionMatch(BaseModel):
    user_id: str = Field(description="ID of the speaker")
    quote: str = Field(description="The exact quote from the conversation, leave empty string if not relevant")
    suggestion: str = Field(
        description="A faithful paraphrasing of the explicit suggestion/idea in one sentence so that it is standalone, without adding interpretations beyond the original text. Leave empty string if not relevant or if the utterance only contains questions, observations, or entity mentions without explicit proposals. If a single utterance contains several unrelated ideas/suggestions, include a separate paraphrased sentence for each."
    )

class SuggestionBatch(BaseModel):
    suggestions: List[SuggestionMatch] = Field(
        description="List of suggestion classifications, one for each utterance in the current window being evaluated. MUST maintain same order as input utterances.",
        min_items=1
    )
    
    @field_validator('suggestions')
    def validate_list_length(cls, v, info):
        expected_length = info.context.get('expected_length') if info.context else None
        if expected_length is not None and len(v) != expected_length:
            raise ValueError(f"Expected {expected_length} classifications, got {len(v)}")
        return v
\end{lstlisting}   
\end{mdframed}

\section{Missing, Viable Suggestions}\label{appendix:student-suggestions}
\subsection{Manual Analysis Motivation}\label{appendix:manual}
The decision to manually analyze the suggestions at this point was due to time restraints. 
It is certainly possible to do this using LLMs and automate this process in the future, however, developing prompts and manually evaluating the LLM outputs was unfortunately not possible within the time constraints of this work.

Also, this illustrates how LLMs are not always faster or better at these sorts of tasks and need not always be involved. That being said, in the future we'd suggest \textit{partially} automating the task by leveraging LLMs to classify suggestions covered by the recommendations and allow for scaling the analysis to more suggestions. The process of deciding which of these distinct recommendations are actually viable requires a lot of nuance and understanding of feasibility and assembly context is best left for a human (at least for now). Moreover, we learned a lot in doing this and it could potentially be a valuable activity in and of itself for the delegates!

\subsection{Manual Analysis Methodology Example}\label{appendix:manualex}
\textbf{Example: Food \& Dining} below contains the details of how we identified the promising distinctive suggestions for that topic. We took the same approach for the rest of the suggestions, the results of which are summarized in \Cref{tab:distinct} and the full list can be found in \Cref{appendix:student-suggestions}.

\begin{table*}[h!]
    \centering
    \begin{tabular}{l|ccc}
        Topic & \# Suggestions $\rightarrow$& \# Distinct $\rightarrow$&\# Unique \\\hline
        Food \& Dining & 15 & 8 & 3\\
        Campus Greening &1&1&1\\
        Innovation \& Entrepreneurship&2&1&1\\
        Information \& Education & 8&3&1\\
        Waste Management&5&4&3\\
        Transportation&4&3&3\\
        Renewable Energy \& Emissions Reduction&12&1&1\\
        Divestment&7&0&0\\\hline
        Totals&54&21&13
    \end{tabular}
    \caption{This table summarizes the process of identifying which of the 54 concrete suggestions are distinct from the recommendations to create the final list of 13 suggestions that may be viable recommendations. To interpret this table, we see for example that while there were 12 suggestions pertaining to renewable energy and emissions reduction, only one of these was actually distinct from the final recommendations and was thus included in the final list. On the other hand, there were 8 suggestions on information and education, only 3 of which were distinct from the recommendations. However, there was only 1 unique idea among these 3, so only the best suggestion between them was chosen for the final list. }
    \label{tab:distinct}
\end{table*}

\subsubsection*{Example: Food \& Dining \normalfont{(15 suggestions)}} 
\begin{mdframed}
Relevant existing recommendations:
\begin{enumerate}[align=left]
    \item[- Rec 1:] ``Waive meal plan requirements for all students in all dorms, and make unused food for the day available to students who need it at a reduced price."
    \item[- Rec 3:] ``Reduce food waste through increased composting around campus:
1) Have compost bins in residence hall kitchens and kitchens around campus in every building
2) Increase signage and have posters near and around compost bins showing people what can be composted and why it is important"
\item[- Rec 45:] ``We need greater support for [Office of Sustainability]'s food waste fighters program to increase accessibility to composting around campus via adding compost bins, especially in/around dorms. This also includes putting informational posters on how to recycle/compost at every waste disposal location and common space (like elevators and lounges) to increase waste management education."
\end{enumerate}

We considered each of the 15 suggestions and noted whether they were covered by any of the above recommendations. This was the case for 7 of the 15 (notably, six of these pertained solely to Rec 1). For the remaining 8 suggestions distinct from the recommendations, we noted that several were redundant with respect to each other, so in these cases we chose the most comprehensive yet concrete one as the unique representative. To give an idea of how redundant suggestions were grouped and which was chosen for the final list, the list below contains the 8 suggestions distinct from the recommendations. The numbered bullet points were the chosen unique suggestions, and the sub-bullet points are corresponding redundant suggestions.

\begin{enumerate}
    \item ``I propose that MIT should implement the use of reusable dining ware in campus dining halls throughout the school year instead of relying on unsustainable paper utensils."
    \begin{itemize}
        \item ``I propose that MIT should eliminate the use of paper plates when dishwashers are not in use."
        \item ``I suggest that MIT should begin with small-scale sustainable catering initiatives, such as using reusable plates, and then gradually expand these practices."
    \end{itemize}
    \item ``I propose that MIT should address the food security issue by expanding kitchen facilities and involving the MIT Food Security Action Team and MIT facilities to provide more access to ingredients for students."
    \begin{itemize}
        \item ``I propose that MIT should create a campus market offering cheap, accessible groceries to help students cook for themselves and reduce food waste in the dining halls."
        \item ``I suggest that MIT should increase the maintenance of cooking areas to improve their usability."
    \end{itemize}
    \item ``I propose that MIT should adopt a serving system with portion size limitations in its dining halls to reduce food waste."
\end{enumerate}

Lastly, there was one vague suggestion that would be covered by the combination of the above: ``I suggest that MIT should update the dining halls overall."
\end{mdframed}

Out of the 54 concrete suggestions identified, only 21 of them were distinct from the recommendations. Furthermore, of the 21, many were not distinct from each other (i.e. were redundant) and as a result, 13 suggestions were selected for the final list of distinct suggestions that could feasibly have become recommendations. These results demonstrate that while the recommendations are fairly comprehensive of the suggestion space, there are still a few that slipped through the cracks. 

One plausible explanation is that perhaps these suggestions were voiced at times in which drafting was not the main focus and were therefore forgotten when it was time to write down the recommendations. However, we found that these 13 missing suggestions were raised across all four days equally: three during the pre-assembly conversations, three on Day 1, three on Day 2, and four on Day 3. 
This suggests that time or forgetfulness cannot fully explain the reason for these suggestions to have been missed, but may still play a factor for the pre-conversation or Day 1 suggestions. Furthermore, of those from Day 2, two were said during the same drafting breakout and pertained to the dining and food waste topic. As seen above, three recommendations were written on this subject, suggesting that there may have been an element of fatigue or hesitancy in writing too many recommendations on the same topic. 
Another consideration in trying to understand why some suggestions did not persist could be related to social factors. We searched for whether there might be any patterns in which delegates voiced these suggestions. We found that the 13 suggestions actually came from 11 unique delegates, indicating that no particular delegate was systematically ignored. 

\subsection{List of 13 Distinct, Viable Suggestions}
The categorized list below contains all 13 missing viable suggestions along with the original transcript quote from the delegate speaker.
\paragraph{Food \& Dining}
\begin{itemize}
    \item I propose that MIT should address the food security issue by expanding kitchen facilities and involving the MIT Food Security Action Team and MIT facilities to provide more access to ingredients for students.
    \begin{itemize}
        \item Transcript quote: ``\textit{There is a food security issue... if more students could be doing the CFY thing they're also going to need, you know, more access to ingredients. So MIT FSAT and if we're going to expand the kitchens, even MIT facilities like involved...}" --Delegate 18, Day 2 Draft Initial Recommendations
    \end{itemize}
    \item I propose that MIT should adopt a serving system with portion size limitations in its dining halls to reduce food waste.
    \begin{itemize}
        \item Transcript quote: ``\textit{Do you guys think that in the dining halls having sort of a serve yourself system is contributing to food waste? Because I feel like sometimes, you know, people just take more than they can eat just so they don't have to get up a second time. But I feel like if we had sort of like the serving system, like, I know [dorm name] has more of a serving system. Portion sizes could be limited because, I mean, like, you can always get up and get a second plate, you know? Do you think that would be worth recommending?}" --Delegate 15, Day 2 Draft Initial Recommendations
    \end{itemize}
    \item I propose that MIT should implement the use of reusable dining ware in campus dining halls throughout the school year instead of relying on unsustainable paper utensils.
    \begin{itemize}
        \item Transcript quote: ``\textit{What I imagine is that we'll definitely kind of not rely that much on, on paper utensils, for example, in dining halls these days I've been seeing a lot of paper plates and forks used while the dishwasher or something are not used when there's less people on campus. And then recently they switched back to reusable plates and cups and stuff. So I think that's something we should try to make use of more or less throughout the entire school year to rely on those instead of like those unsustainable ones.}" --Delegate 3, Pre-Assembly Conversation
    \end{itemize}
\end{itemize}

\paragraph{Transportation}
\begin{itemize}
    \item I suggest that MIT implement a parking fee of \$600 per semester for limited campus parking spaces to discourage driving and to redirect the funds towards sustainable initiatives.
    \begin{itemize}
        \item Transcript quote: ``\textit{Maybe something else the university could do is to say that with a very limited number of parking spaces on campus, you have to pay \$600 to have a parking space for the semester. So to kind of further discourage people from driving on campus that way, and then that money could be used to further fund sustainable efforts.}" --Delegate 14, Day 1 Breakout 3
    \end{itemize}
    \item I suggest that MIT should implement stipends for electric vehicle owners to incentivize the adoption of environmentally friendly transportation.
    \begin{itemize}
        \item Transcript quote: ``\textit{My personal beliefs is centered more towards research and, like, finding economical situations to the. Where it's like, beneficial to want to switch. Yeah, I know one policy I did agree with recently is I don't know if it's Florida specific, but they do give out. It probably has to do with Elon Musk, but they do give out stipends if you have electric vehicles, which I do, like, think is a really good incentive because it just improves access and it allows people to want to choose a more environmental option.}" --Delegate 11, Day 1 Discuss Pre-reads
    \end{itemize}
        \item I suggest that MIT reflect on the necessity of travel for conferences and consider organizing some events online to reduce unnecessary travel.
    \begin{itemize}
        \item Transcript quote: ``\textit{What kind of travels are necessary. Like, especially during [January], a lot of people are going on global teaching lab or spending two weeks here and there. Like, and I know people in Argentina and Tenerife and everything. And I get that, like local experiences are really useful, but maybe there could be kind of a reflection behind them or when it comes to inviting like teachers and organizing conferences, maybe switching some of them like online.}" --Delegate 10, Pre-Assembly Conversation
    \end{itemize}
\end{itemize}
\paragraph{Renewable Energy \& Emissions Reduction}
\begin{itemize}
    \item I suggest that MIT implement geothermal stratum machines using the pre-existing cool loop on campus.
    \begin{itemize}
        \item Transcript quote: ``\textit{Like the geothermal stratum machines that we were talking about can be implemented using the pre existing cool loop that's on it's campus}" --Delegate 17, Day 1 Breakout 3
    \end{itemize}
\end{itemize}
\paragraph{Innovation \& Entrepreneurship}
\begin{itemize}
    \item I propose that MIT should implement a 'shark tank' style initiative to foster and fund innovative projects.
    \begin{itemize}
        \item Transcript quote: ``\textit{I'll put a shark tank on board too.}" --Delegate 18, Day 3 Info Fair Debrief
    \end{itemize}
\end{itemize}
\paragraph{Waste Management}
\begin{itemize}
    \item I suggest that MIT should organize a trash picking up competition modeled after a Rex event as a sustainability initiative.
    \begin{itemize}
        \item Transcript quote: ``\textit{What about like a Rex event? Like sustainability related? Like a. I don't know, like a trash picking up competition?}" --Delegate 7, Day 3 Slap Takes
    \end{itemize}
    \item I propose that MIT should consider adopting smaller trash receptacles to reduce waste generation by limiting the available container size.
    \begin{itemize}
        \item Transcript quote: ``\textit{[Moderator name] earlier mentioned this... she was saying that like back when she and her husband lived in Switzerland that they were given a smaller trash bag, right? And when they're given a smaller trash bag, they generate less trash. So I think it's also a matter of perception because here we get three. Okay, at least in California, I don't know, but we get three huge trash bins, one for like every single week. We're generating like what we're given the, the possibility of generating a huge bin of regular trash, a huge bin of like natural trash... Ironically, the recycling bins smallest one. Right. Like that's crazy. But I think it's also about like people perception of the world and seeing that like, because we're given room to actually create that much waste, we are okay with generating that much waste. And it's the same thing here. It's like whenever, you know, my suitemates and I, we fill up the trash can, we would think to ourselves, okay, it's time to maybe slow down.}" --Delegate 19, Day 2 Breakout 1
    \end{itemize}
    \item I suggest that MIT should prioritize addressing the waste and harmful materials generated by research activities to reduce environmental impact on campus.
    \begin{itemize}
        \item Transcript quote: ``\textit{One of the biggest, I guess like, criticisms of MIT that I've heard is that there's a lot of waste generated by the research being done and, or like some harmful material being put out...}" --Delegate 13, Pre-Assembly Conversation
    \end{itemize}
\end{itemize}
\paragraph{Campus Greening}
\begin{itemize}
    \item I propose that MIT should support a student group initiative to raise money for planting trees.
    \begin{itemize}
        \item Transcript quote: ``\textit{an MIT student group raising money for planting trees.}'' --Delegate 14, Day 3 Slap Takes
    \end{itemize}
\end{itemize}
\paragraph{Information \& Education}
\begin{itemize}
    \item I suggest that MIT should implement Tik tok challenges.
    \begin{itemize}
     \item Transcript quote: ``\textit{Tik tok challenges.}" --Delegate 16, Day 3 Info Fair Debrief
    \end{itemize}
\end{itemize}

\section{Delegate Profile Prompt Iteration \& Evaluation}\label{appendix:profileprompts}
Similar to the systematic prompt engineering in \Cref{appendix:promptengineering}, we iterated on the delegate profiles prompt four times based on inspection of false positives and negatives. For evaluation, we again focused on the drafting recommendations breakout sessions due to the abundance of mentions of recommendations. In each round, we read through the conversations entirely to manually inspect outputs for any false positives or negatives to inform what prompt edits were necessary. 

The initial prompt (\Cref{appendix:evidencev1}) often over-generalized topic relevance, assuming that all conversations generally about sustainability were relevant to the specific recommendation at hand and therefore classifying almost everything as relevant. Moreover, the model would mark short utterances as evidence for a specific recommendation (e.g. ``I think that's a good proposal") when there isn't enough context to determine what proposal they're referring to. To mitigate these issues, we revised the prompt to more clearly distinguish between the overarching assembly context and the specific statement topic, require more explicit evidence before categorizing something as relevant, especially for short or ambiguous utterances. Lastly, we also added more explicit criteria for what utterances are not relevant (e.g. general opinions about sustainability without specific connection to the statement or questions seeking clarification without expressing an opinion).

The second prompt version (\Cref{appendix:evidencev2}) was an improvement, but there was a confounding issue wherein the model failed to clearly distinguish between the context and the current utterance being evaluated, leading to confusion of speakers and which utterance to evaluate.
To combat this in the third version (\Cref{appendix:evidencev3}), we added in speaker names more clearly, emphasized to evaluate only the current speaker turn, and required the model to quote specific words in the rationale. 
With this version outputs, in a similar fashion to the suggestions prompt evaluation, we used Claude 3.7 Sonnet to analyze a larger set of \texttt{o3-mini} outputs at this point and calculate a true versus false positive count on all the breakouts pertaining to one recommendations statement. The model estimated that there were only 5\% false positives, but there were a few lingering issues (full model output is in \Cref{appendix:claude-analysis-profiles}):
\begin{itemize}
    \item The model is over-relying on keyword matching (like "divest" or "fossil fuel") where the topic is mentioned but no meaningful stance or perspective is shared.
    \item The rationales aren't explaining how the utterances reveal the speaker's stance beyond just noting the presence of keywords.
\end{itemize}

To give an example of the above issue, one delegate said, ``Maybe about like the divestment stuff. Did you write anything down?" and it was classified as relevant with the LLM rationale saying ``Explicitly mentions 'divestment stuff', directly referencing the proposal to divest. This is a clear signal of discussion regarding the divestment recommendation." However, this is clearly not a meaningful statement on divestment from this delegate.

To address this, we revised the final prompt (\Cref{appendix:evidencev4}) to focus on the delegate's expressed stance with examples of different types of stances (positive, negative, neutral, uncertain). In addition, we clarified that questions should only count if they imply a position (e.g. uncertainty, skepticism, support, etc). We instructed the model to cite relevant conversation context in the rationale when needed to understand the full stance and not rely on keywords. During manual inspection, we found that asking the model to evaluate multiple speaker turns at a time was more effective in both reducing false positives and false negatives because the model was given more context and required to analyze it. As a bonus, processing the transcripts in batches of 10 instead of line-by-line was much cheaper and faster to run! Some examples of false negatives now classified correctly by the final prompt are in \Cref{appendix:falseneg}.

\section{Delegate Profiles Prompt Evaluation}\label{appendix:delprofeval}

\subsection{With Claude}\label{appendix:claude-analysis-profiles}
This analyzes the output of the prompt in \Cref{appendix:evidencev3}: 
\begin{mdframed}
Let's analyze the dataset to determine the proportion of true positives versus false positives:

First, I'll review each entry to determine if it's correctly classified as relevant to the statement "Divest the endowment from fossil fuels within the next 5 years."

I'll classify each utterance as:
- True Positive (TP): Correctly identified as relevant to divestment from fossil fuels
- False Positive (FP): Incorrectly identified as relevant when it's not about divestment

Starting my analysis:

1. Delegate 5: "Like if we could just like completely divest."
   This is clearly about divestment and mentions it explicitly - TP

2. Delegate 5: "Like practice what you preach. Be like we are completely divested from."
   This is explicitly about divestment - TP

3. Delegate 12: "Google. Gemini says according to information available, MIT has not publicly disclosed the exact percentage that's endowment directly tied to fossil fuel companies."
   This is about MIT's endowment and fossil fuel companies - TP

4. Delegate 12: "Yeah, I definitely think the website we were looking at wasn't updated because this is also a new group called MIT Divest."
   Mentions MIT Divest group - TP

5. Delegate 12: "2021. Never mind. Maybe they're like dismantled."
   Referring to the MIT Divest group - TP

6. Delegate 12: "Maybe about like the divestment stuff. Did you write anything down?"
   Explicitly mentions divestment - TP

7. Delegate 12: "Yeah. Or like increased transparency on divestment or like the fossil fuel endowment."
   Explicitly mentions divestment and fossil fuel endowment - TP

8. Delegate 12: "Divest."
   Single word but clearly on topic - TP

9. Delegate 12: "You write divest now."
   Explicitly about divestment - TP

10. Delegate 12: "But technically we're not like saying we should divest. We're just saying increase transparency. Right?"
    Explicitly discusses divestment vs transparency - TP

Continuing through the list:

All of these utterances appear to be explicitly discussing divestment from fossil fuels, MIT's endowment investments in fossil fuels, or directly related topics. They contain key terms like "divest," "fossil fuel companies," "endowment," etc.

The only borderline cases might be very short utterances like "Divest." or responses that don't repeat the term "divest" but are clearly responding to a comment about divestment.

Let me check for any possible false positives:

1. Delegate 16: "Oh yeah, I did three years ago." - This could be a false positive if the context doesn't clearly establish that "divestment" is what's being discussed. However, the rationale says it's a response to a divestment remark, which suggests it's likely a true positive.

2. Delegate 16: "I don't know. How do I write that? What's the recommendation?" - This is potentially a false positive if there's ambiguity about what recommendation they're discussing. But if the conversation is clearly about divestment, it's likely a true positive.

3. Delegate 14: "Do it like 30\%. I get that that's a large chunk. Like 1\% is not, it's nothing." - Could be a false positive if the context isn't clearly about divestment percentages, but it sounds like it likely is.

4. Delegate 4: "Like I don't think financial." - This is a very short fragment that on its own doesn't seem to be about divestment, but it's categorized as relevant because it's a response to a question about divestment and financial aid. It's borderline.

5. Delegate 7: "They have the power to do so. I don't know how we're going to put that recommendation." - Without more context, it's not explicit whether this is specifically about divestment, but it's categorized as such.

After reviewing all entries, I find that the vast majority appear to be true positives - they explicitly mention divestment, fossil fuels, or the endowment, or they are direct responses to such comments. There are very few potential false positives where the relevance is questionable.

Let me count:
- Total utterances: 60

- Clear true positives: approximately 57

- Potential false positives: approximately 3 (but these are borderline cases)

This gives us approximate proportions:

- True positives: 57/60 = 0.95 (95\%)

- False positives: 3/60 = 0.05 (5\%)

The revised prompt appears to be working quite well for this statement, with approximately 95\% of the identified utterances being true positives. The model is correctly identifying utterances that explicitly mention "divest," "fossil fuels," and related concepts, as well as direct responses to such comments that provide meaningful context about speakers' perspectives.

However, without more context for those borderline cases, it's difficult to be certain whether they are truly false positives.
\end{mdframed}

\subsection{Examples of False Negatives}\label{appendix:falseneg}
Below is an excerpt of the results comparing from \Cref{appendix:evidencev4} (which evaluates ten utterances at a time). We see that this final prompt version correctly identified these three utterances as relevant to the statement, whereas the previous prompt (\ref{appendix:evidencev3}) missed these (i.e. false negatives). 
\begin{mdframed}
\begin{lstlisting}
=== Statement: MIT should implement an educational program similar to [university program] for the purpose of educating youth about low-cost sustainable practices and climate science. The program should be no-cost for participants, and prioritize those from low-income communities or those where climate science is not included in public school district curriculum. When possible, MIT student educators should be matched to programs in communities similar to their own background, to improve communicability and understanding of local climate challenges. ===

Conversation: day2\_DraftInitialRecs.csv
-------------------------------------
 Delegate 3 (delegate):                                                                
 - Utterance: I talked about, like, just implementing programs about, like, sustainability to elementary or secondary schools.
   Rationale: The speaker explicitly mentions 'implementing programs about, like, sustainability to elementary or secondary schools,' which directly connects to the idea of educating youth about sustainability as proposed in the statement.
   Confidence: 0.9                                                               
-------------------------------------
 Delegate 3 (delegate):                                                                
 - Utterance: I wrote one on like the programs for, I guess, sustainability, like for secondary and elementary school students. And then I'm writing another one on like improving the public transportation for MIT. Like the tech shuttles. Because I. I know the current ones are like, really bad and no one, like, wants to use it because it's really inaccurate. And then like, sometimes you have to wait for an hour for like the bus.
   Rationale: The speaker explicitly states: 'I wrote one on like the programs for, I guess, sustainability, like for secondary and elementary school students,' directly addressing an educational program for youth-a key element of the proposal.
   Confidence: 0.8                                                               
-------------------------------------
 Delegate 3 (delegate):                                                                
 - Utterance: Like rsi, but for, like, sustainability.                           
   Rationale: The speaker compares the idea to 'rsi, but for, like, sustainability,' which implies an educational program for youth focused on sustainability-directly aligning with the statement's proposal.
   Confidence: 0.9                                  
\end{lstlisting}
\end{mdframed}

\section{Delegate Profiles Prompts}\label{appendix:delegateprofileprompts}

\subsection{Initial Prompts}\label{appendix:evidencev1}
Below are the first prompts we used to create the delegate profiles from the transcripts using OpenAI's \texttt{o3-mini} model:

\begin{mdframed}[frametitle={System Instructions:},
    frametitlefont=\bfseries,
    innertopmargin=1em
]
\begin{lstlisting}
Your task is to find quotes from a conversation transcript, which is broadly on the topic: {topic}.

Concretely, you will be given a specific statement and an excerpt of the transcript. You will use the below criteria to decide if the current utterance is relevant to the given statement.
    Category 1. **MOST IMPORTANT: is the current utterance relevant to understanding the speaker's opinions/perspectives on this statement in particular?** This may include:
        a. The utterance *explicitly* mentions key concepts or actions present in or *specifically connected* to the statement
        b. The utterance discusses problems that the statement aims to solve or reframes those problems
        c. The utterance proposes relevant solutions that align with the statement's goals or, conversely, proposes solutions that are in conflict or contrasts with the statement's goals
        d. The utterance evaluates or comments on very similar initiatives to those in the statement
        e. The utterance expresses agreement or disagreement with the statement directly
    Category 2. If the current utterance is responding directly to any of the previous utterances (but is not relevant standalone) and the above criteria apply to the utterance it responds directly to, then this utterance is relevant.
    Category 3. Otherwise, there is no strong relevance between the utterance and the statement. Utterances related to the broad conversation topic but not specifically to the statement are considered not relevant.
\end{lstlisting}
\end{mdframed}

In the prompt, the \texttt{\{statement\}} is the text of the recommendation, \texttt{\{curr\_utterance\}} is one speaker turn from the transcript to evaluate, and \texttt{\{context\}} are the 10 speaker turns in the transcript preceding the current utterance. 

\begin{mdframed}[frametitle={Prompt:},
    frametitlefont=\bfseries,
    innertopmargin=1em
]
\begin{lstlisting}
## **Statement:** 
"{statement}"

## **Previous context:**
{context}

## **Current utterance:**
{curr_utterance}

## **Instructions** 

**Relevance definition: an utterance is relevant to a statement if it is helpful in understanding the speaker's opinions/perspectives on this statement in particular.**
To determine if the current utterance is relevant to this statement, follow these steps:

Step 1. Using the criteria outlined in the system prompt, determine which of the three categories are applicable. Indicate the category number.
Step 2. If category 1, provide a rationale for your decision.
Step 3. If it is category 2 (the current utterance is responding directly to a previous utterance which is directly relevant), give the **exact quote** of the utterance the current one responds to.
Step 4. If it is category 2, provide a rationale for why this utterance is relevant to the previous, and why the previous is in category 1.
Step 5. Indicate your confidence level on the relevance, where 0 is completely irrelevant and 1 is extremely relevant.
Step 6. If the current utterance is not relevant (category 3), leave the above fields blank and give a score of 0.0.

You must return a JSON object matching this schema:
{schema_string}
\end{lstlisting}
\end{mdframed}

The \texttt{schema\_string} specifies the output formatting, which is a JSON  containing the classification of current utterance from the transcript (category 1 is relevant, category 2 is responding substantively to a relevant prior speaker turn, and category 3 is irrelevant), a rationale explaining why, and a confidence score. The schema is defined by the Pydantic model below:

\begin{mdframed}[frametitle={Model Schema:},
    frametitlefont=\bfseries,
    innertopmargin=1em
]
\begin{lstlisting}[language=Python,showstringspaces=false]
class EvidenceMatchV1(BaseModel):
    category: str = Field(description="1, 2, or 3", pattern="^[123]$")
    llm_rationale: str = Field(description="(if category 1, otherwise leave empty '') Explanation of why this utterance specifically helps understand the current speaker's view on the statement, otherwise empty string")
    response_quote: str = Field(description="(if category 2, otherwise leave empty '') The exact quote of the previous utterance for which the current utterance responds to (only for category 2)")
    llm_response_rationale: str = Field(description="(if category 2, otherwise leave empty '') Rationale for why this utterance, taken along with the previous one it responds to, helps understand the current speaker's view on the statement")
    confidence_score: float = Field(
        description="Confidence score between 0 and 1 indicating relevance",
        ge=0,
        le=1
    )    
\end{lstlisting}   
\end{mdframed}

\subsection{Second Prompt Iteration}\label{appendix:evidencev2}
Below are the second iteration of prompts we used to create the delegate profiles from the transcripts using \texttt{o3-mini}:

\begin{mdframed}[frametitle={System Instructions:},
    frametitlefont=\bfseries,
    innertopmargin=1em
]
\begin{lstlisting}
You are analyzing transcripts from an MIT deliberative assembly about sustainability. Your task is to find quotes that are relevant to a specific statement/proposal.

While ALL conversations relate to the broad topic of campus sustainability, you must ONLY identify utterances specifically addressing the narrower STATEMENT topic provided. The statement represents a specific proposal/recommendation, not the general topic of sustainability.

Use these strict criteria to determine relevance:

Category 1. **DIRECT RELEVANCE: The utterance provides clear evidence of the speaker's opinions/perspectives on the SPECIFIC statement.**
    a. The utterance *explicitly* mentions key terms from the statement OR *directly* discusses the specific actions proposed
    b. The utterance clearly evaluates the specific proposal in the statement (not just any sustainability initiative)
    c. The utterance discusses specific problems that the statement aims to solve or specific solutions aligned with the statement
    d. The utterance expresses a clear position (agreement/disagreement) specifically about the statement

Category 2. **CONTEXTUAL RELEVANCE: The utterance directly responds to a Category 1 utterance and adds context to understand opinions.**
    For this category, BOTH these conditions must be met:
    - The current utterance directly and substantively responds to a Category 1 utterance
    - The combination reveals something meaningful about the speaker's view on the statement
    - Simple acknowledgments or general follow-ups do NOT qualify

Category 3. **NOT RELEVANT**
    - Comments about other sustainability initiatives not mentioned in the statement
    - General opinions about sustainability without specific connection to the statement
    - Questions seeking clarification without expressing an opinion on the statement
    - Brief responses that don't reveal any opinion on the statement
    - Discussion of topics that appear similar but aren't directly about the statement content

IMPORTANT NOTES ON SHORT UTTERANCES:
Short utterances like "I agree" or "That's good" are ONLY relevant if the immediate preceding context makes it UNAMBIGUOUS that they are discussing the specific statement topic.

IMPORTANT NOTES ON QUESTIONS:
Questions are NOT evidence of a speaker's opinion unless they express an implicit position (e.g., "Wouldn't composting cost too much?") or contain substantive reasoning related to the statement. Simple clarifying questions should be marked Category 3.
\end{lstlisting}
\end{mdframed}

In the prompt, the \texttt{\{statement\}} is the text of the recommendation, \texttt{\{curr\_utterance\}} is one speaker turn from the transcript to evaluate, and \texttt{\{context\}} are the 10 speaker turns in the transcript preceding the current utterance. 

\begin{mdframed}[frametitle={Prompt:},
    frametitlefont=\bfseries,
    innertopmargin=1em
]
\begin{lstlisting}
## **Statement:** 
{statement}

## **Previous context:**
{context}

## **Current utterance:**
{curr_utterance}

## **Instructions** 

**Relevance definition: an utterance is relevant to a statement if it is helpful in understanding the speaker's opinions/perspectives on this statement in particular.**

To determine if the current utterance is relevant to this statement, follow these steps:

Step 0: Verify that the context or current utterance explicitly mentions the statement topic or its key elements. If there is no clear evidence that they are discussing the specific statement topic, default to Category 3 with confidence 0.0.

Step 1: Using the criteria outlined in the system prompt, determine which of the three categories are applicable. Indicate the category number.

Step 2: If category 1, provide a rationale explaining EXACTLY how this utterance reveals the speaker's perspective on the SPECIFIC statement (not just general sustainability).

Step 3: If it is category 2 (the current utterance is responding directly to a previous utterance which is directly relevant), give the **exact quote** of the utterance the current one responds to.

Step 4: If it is category 2, provide a rationale explaining how this response, combined with the previous utterance, reveals the speaker's perspective on the SPECIFIC statement.

Step 5: Indicate your confidence level on the relevance, where 0 is completely irrelevant and 1 is extremely relevant. Be conservative - only use high scores (>0.7) when evidence is very clear.

Step 6: If the current utterance is not relevant (category 3), leave the above fields blank and give a score of 0.0.

Remember:
- DO NOT assume conversations are about the statement topic without clear evidence
- DO NOT categorize general sustainability discussions as relevant
- Brief or ambiguous responses require stronger contextual evidence to be relevant

You must return a JSON object matching this schema:
{schema_string}
\end{lstlisting}
\end{mdframed}

The \texttt{schema\_string} specifies the output formatting, which is a JSON  containing the classification of current utterance from the transcript (category 1 is relevant, category 2 is responding substantively to a relevant prior speaker turn, and category 3 is irrelevant), whether the topic was explicitly mentioned in the utterance, a rationale justifying ir/relevance, and a confidence score. If category 2, the LLM must also include the quote of the speaker turn that the current one responds to. The schema is defined by the Pydantic model below:

\begin{mdframed}[frametitle={Model Schema:},
    frametitlefont=\bfseries,
    innertopmargin=1em
]
\begin{lstlisting}[language=Python,showstringspaces=false]
class EvidenceMatchV2(BaseModel):
    category: str = Field(
        description="1 (direct relevance), 2 (contextual relevance), or 3 (not relevant)",
        pattern="^[123]$"
    )
    topic_mentioned: bool = Field(
        description="Whether the statement topic or its key elements are explicitly mentioned in the context or current utterance"
    )
    llm_rationale: str = Field(
        description="(if category 1, otherwise leave empty '') Explain EXACTLY how this utterance reveals the speaker's perspective on the SPECIFIC statement, citing exact keywords or concepts from the statement that appear"
    )
    response_quote: str = Field(
        description="(if category 2, otherwise leave empty '') The exact quote of the previous utterance that this utterance responds to, which must itself be directly relevant to the statement (category 1)"
    )
    llm_response_rationale: str = Field(
        description="(if category 2, otherwise leave empty '') Explanation of how this response, combined with the previous utterance, reveals something meaningful about the speaker's view on the SPECIFIC statement "
    )
    irrelevance_reason: str = Field(
        description="(if category 3, otherwise leave empty '') Brief explanation of why this utterance is not relevant to the specific statement (e.g., discusses different topic, too general, etc.)"
    )
    confidence_score: float = Field(
        description="Confidence score between 0 and 1 indicating relevance. Be conservative - only use high scores (>0.7) when evidence is very clear.",
        ge=0,
        le=1
    )
\end{lstlisting}   
\end{mdframed}

\subsection{Third Prompt Iteration}\label{appendix:evidencev3}
Below are the third iteration of prompts we used to create the delegate profiles from the transcripts using \texttt{o3-mini}:

\begin{mdframed}[frametitle={System Instructions:},
    frametitlefont=\bfseries,
    innertopmargin=1em
]
\begin{lstlisting}
You are analyzing transcripts from an MIT deliberative assembly about sustainability. Your task is to find quotes that are relevant to a specific statement/proposal.

While ALL conversations relate to the broad topic of campus sustainability, you must ONLY identify utterances specifically addressing the narrower STATEMENT topic provided. The statement represents a specific proposal/recommendation, not the general topic of sustainability.

IMPORTANT: You must ONLY evaluate the CURRENT UTTERANCE provided, not anything in the previous context. The previous context is only provided to help you understand what the current utterance is responding to.

Use these strict criteria to determine relevance:

Category 1. **DIRECT RELEVANCE: The CURRENT utterance provides clear evidence of the speaker's opinions/perspectives on the SPECIFIC statement.**
    a. The CURRENT utterance *explicitly* mentions key terms from the statement OR *directly* discusses the specific actions proposed
    b. The CURRENT utterance clearly evaluates the specific proposal in the statement (not just any sustainability initiative)
    c. The CURRENT utterance discusses specific problems that the statement aims to solve or specific solutions aligned with the statement
    d. The CURRENT utterance expresses a clear position (agreement/disagreement) specifically about the statement

Category 2. **CONTEXTUAL RELEVANCE: The CURRENT utterance directly responds to a previous utterance that has Category 1 relevance.**
    For this category, BOTH these conditions must be met:
    - The CURRENT utterance directly and substantively responds to a Category 1 utterance in the previous context
    - This combination reveals something meaningful about the CURRENT speaker's view on the statement
    - Simple acknowledgments or general follow-ups do NOT qualify

Category 3. **NOT RELEVANT: The CURRENT utterance does not meet Category 1 or 2 criteria.**
    - CURRENT utterance has no direct mention of statement topics
    - CURRENT utterance contains general comments about sustainability unrelated to the statement
    - CURRENT utterance is a clarifying question without expressing an opinion
    - CURRENT utterance is too brief or ambiguous to determine relevance to statement

Before categorizing, check if the CURRENT utterance (not the context) contains specific references to the statement topic.
\end{lstlisting}
\end{mdframed}

In the prompt, the \texttt{\{statement\}} is the text of the recommendation, \texttt{\{curr\_utterance\}} is one speaker turn from the transcript to evaluate, \texttt{\{curr\_speaker\}} is the name of that speaker, and \texttt{\{context\}} are the 10 speaker turns in the transcript preceding the current utterance. 

\begin{mdframed}[frametitle={Prompt:},
    frametitlefont=\bfseries,
    innertopmargin=1em
]
\begin{lstlisting}
## **Statement:** 
{statement}

## **Previous context:**
{context}

## **Current utterance to evaluate (ONLY THIS):**
Speaker: {curr_speaker}
Utterance: "{curr_utterance}"

## **Instructions** 

**Relevance definition: an utterance is relevant to a statement if it helps understand the speaker's opinions/perspectives on this statement in particular.**

To determine if THIS SPECIFIC CURRENT utterance is relevant to the statement, follow these steps:

Step 0: First, CONFIRM you are only evaluating the current utterance by Speaker: {curr_speaker}, not anything in the previous context.

Step 1: Verify that the CURRENT utterance EXPLICITLY mentions the statement topic or its key elements. If the current utterance does not contain clear references to the statement topic, it is likely Category 3.

Step 2: Using the criteria outlined in the system prompt, determine which of the three categories are applicable. Indicate the category number.

Step 3: If category 1, provide a rationale explaining EXACTLY how THIS CURRENT utterance reveals the speaker's perspective on the SPECIFIC statement by quoting specific words or phrases FROM THE CURRENT UTTERANCE.

Step 4: If it is category 2, give the **exact quote** from the previous context that the current utterance responds to.

Step 5: If it is category 2, provide a rationale explaining how this response, combined with the previous utterance, reveals the speaker's perspective on the SPECIFIC statement.

Step 6: Indicate your confidence level on the relevance, where 0 is completely irrelevant and 1 is extremely relevant. Be conservative - only use high scores (>0.7) when evidence is very clear.

Step 7: If the current utterance is not relevant (category 3), provide a brief explanation why.

CRITICAL REMINDER:
- You are ONLY evaluating: Speaker: {curr_speaker}, Utterance: "{curr_utterance}"
- DO NOT evaluate anything from the previous context
- DO NOT assume the current utterance refers to the statement without explicit evidence IN THE CURRENT UTTERANCE
- If in doubt, categorize as Category 3 (not relevant)

You must return a JSON object matching this schema:
{schema_string}
\end{lstlisting}
\end{mdframed}

The \texttt{schema\_string} specifies the output formatting, which is a JSON  containing the classification of current utterance from the transcript (category 1 is relevant, category 2 is responding substantively to a relevant prior speaker turn, and category 3 is irrelevant), whether the topic was explicitly mentioned in the utterance, a rationale justifying ir/relevance, and a confidence score. If category 2, the LLM must also include the quote of the speaker turn that the current one responds to. The schema is defined by the Pydantic model below:

\begin{mdframed}[frametitle={Model Schema:},
    frametitlefont=\bfseries,
    innertopmargin=1em
]
\begin{lstlisting}[language=Python,showstringspaces=false]
class EvidenceMatchV3(BaseModel):
    category: str = Field(
        description="1 (direct relevance), 2 (contextual relevance), or 3 (not relevant)",
        pattern="^[123]$"
    )
    topic_mentioned: bool = Field(
        description="Whether the statement topic or its key elements are EXPLICITLY mentioned in the CURRENT utterance (not in the context)"
    )
    llm_rationale: str = Field(
        description="(if category 1, otherwise leave empty '') Explain EXACTLY how the CURRENT utterance reveals the speaker's perspective, QUOTING SPECIFIC WORDS OR PHRASES from the current utterance only"
    )
    response_quote: str = Field(
        description="(if category 2, otherwise leave empty '') The exact quote FROM THE PREVIOUS CONTEXT that the current utterance directly responds to"
    )
    llm_response_rationale: str = Field(
        description="(if category 2, otherwise leave empty '') Explanation of how the CURRENT utterance, combined with the previous utterance it responds to, reveals the speaker's view on the statement"
    )
    irrelevance_reason: str = Field(
        description="(if category 3, otherwise leave empty '') Brief explanation of why the CURRENT utterance is not relevant to the specific statement"
    )
    confidence_score: float = Field(
        description="Confidence score between 0 and 1 indicating relevance. Be conservative - only use high scores (>0.7) when evidence is very clear.",
        ge=0,
        le=1
    )
\end{lstlisting}   
\end{mdframed}

\subsection{Final Prompts}\label{appendix:evidencev4}
Below are the final iteration of prompts we used to create the delegate profiles from the transcripts using \texttt{o3-mini}:

\begin{mdframed}[frametitle={System Instructions:},
    frametitlefont=\bfseries,
    innertopmargin=1em
]
\begin{lstlisting}
You are analyzing transcripts from an MIT deliberative assembly about sustainability. Your task is to find quotes that are relevant to a specific statement/proposal.

While ALL conversations relate to the broad topic of campus sustainability, you must ONLY identify utterances specifically addressing the narrower STATEMENT topic provided. The statement represents a specific proposal/recommendation, not the general topic of sustainability.

You will be given MULTIPLE UTTERANCES to evaluate at once. For EACH utterance, determine if it is relevant to the statement using these criteria:

Category 1. **DIRECT RELEVANCE: The utterance provides clear evidence of the speaker's opinions/perspectives on the SPECIFIC statement.**
    a. The utterance *explicitly* mentions key terms from the statement OR *directly* discusses the specific actions proposed
    b. The utterance clearly evaluates the specific proposal in the statement (not just any sustainability initiative)
    c. The utterance discusses specific problems that the statement aims to solve or specific solutions aligned with the statement
    d. The utterance expresses a clear position (agreement/disagreement) specifically about the statement

Category 2. **CONTEXTUAL RELEVANCE: The utterance directly responds to a previous utterance that has Category 1 relevance.**
    For this category, BOTH these conditions must be met:
    - The utterance directly and substantively responds to a Category 1 utterance in the previous context
    - This combination reveals something meaningful about the speaker's view on the statement
    - Simple acknowledgments or general follow-ups do NOT qualify

Category 3. **NOT RELEVANT: The utterance does not meet Category 1 or 2 criteria.**
    - Utterance has no direct mention of statement topics
    - Utterance contains general comments about sustainability unrelated to the statement
    - Utterance is a clarifying question without expressing an opinion
    - Utterance is too brief or ambiguous to determine relevance to statement

You must evaluate EACH utterance independently, even if they are related to each other.
\end{lstlisting}
\end{mdframed}

In the prompt, the \texttt{\{statement\}} is the text of the recommendation, \texttt{\{curr\_utterancea\}} contains 10 speaker turns from the transcript to evaluate, and \texttt{\{context\}} are the 10 preceding speaker turns in the transcript. 

\begin{mdframed}[frametitle={Prompt:},
    frametitlefont=\bfseries,
    innertopmargin=1em
]
\begin{lstlisting}
## **Statement:** 
{statement}

## **Previous context:**
{context}

## **Current utterances to evaluate:**
{curr_utterances}

## **Instructions** 

For EACH utterance in the list above:

1. You MUST return classifications in the SAME ORDER as the utterances above, where the first item in your JSON list corresponds to utterance [1], the second to utterance [2], and so on.

2. For each utterance, determine:
   - Category (1, 2, or 3)
   - Whether topic is mentioned explicitly
   - Rationale based on category:
     * Category 1: Explain how this utterance reveals the speaker's perspective
     * Category 2: Include the previous quote being responded to and explain the connection
     * Category 3: Brief explanation of why not relevant
   - Confidence score (0-1, be conservative)

Return a JSON list matching this schema, with one entry per utterance:
{schema_string}

CRITICAL REMINDERS:
- Your output list MUST include ALL input utterances
- Evaluate each utterance independently
- Only consider explicit references to the statement topic
- If in doubt, categorize as Category 3 (not relevant)
\end{lstlisting}
\end{mdframed}

The \texttt{schema\_string} specifies the output formatting, which is a list of JSONs, each with the same fields as in the previous version. The schema is defined by the Pydantic model below, which also has a field validator to ensure that the model returns a valid classification for all 10 speaker turns that were input.

\begin{mdframed}[frametitle={Model Schema:},
    frametitlefont=\bfseries,
    innertopmargin=1em
]
\begin{lstlisting}[language=Python,showstringspaces=false]
class EvidenceMatchBatchV4(BaseModel):
    utterance_classifications: List[EvidenceMatchV3] = Field(
        description="List of classifications, one for each utterance in the current window being evaluated. MUST maintain same order as input utterances.",
        min_items=1  # Allow for partial windows at end of conversation
    )
    
    @field_validator('utterance_classifications')
    def validate_list_length(cls, v, info):
        # Get expected length from context if available
        expected_length = info.context.get('expected_length') if info.context else None
        if expected_length is not None and len(v) != expected_length:
            raise ValueError(f"Expected {expected_length} classifications, got {len(v)}")
        return v
\end{lstlisting}   
\end{mdframed}

\end{document}